\documentclass{aa}  

\usepackage{graphicx}
\usepackage{txfonts}
\usepackage{pdflscape}
\begin{document}

   \title{New models for the evolution of post-asymptotic giant branch stars
     and central stars of planetary nebulae. }
\titlerunning{ New models for the evolution of Post-AGB stars and CSPNe.}
\authorrunning{Marcelo M. Miller Bertolami}

   \author{Marcelo Miguel Miller Bertolami
          \inst{1,2}\fnmsep\thanks{Postdoctoral
fellow of the Alexander von Humboldt Foundation.}
          }

   \institute{Max-Planck-Institut f\"ur Astrophysik,
 Karl-Schwarzschild-Str. 1, 8574, Garching, Germany.\\
              \email{marcelo@mpa-garching.mpg.de}
         \and
           Instituto de Astrof\'isica de La Plata, UNLP-CONICET,
    Paseo del Bosque s/n, 1900 La Plata, Argentina.\\
             \email{mmiller@fcaglp.unlp.edu.ar}
             }

   \date{Received February 29, 1 BC; accepted February 30, 2016}

 
  \abstract {The post-asymptotic giant branch (AGB) phase is arguably
    one of the least understood phases of the evolution of low- and
    intermediate- mass stars. The two grids of models presently
    available are based on outdated micro- and macrophysics and do not
    agree with each other. Studies of the central stars of planetary
    nebulae (CSPNe) and post-AGB stars in different stellar
    populations point to significant discrepancies with the
    theoretical predictions of post-AGB models.}  {We study the
    timescales of post-AGB and CSPNe in the context of our present
    understanding of the micro- and macrophysics of stars. We want to
    assess whether new post-AGB models, based on the latter
    improvements in TP-AGB modeling, can help us to understand the
    discrepancies between observation and theory and within theory
    itself. In addition, we aim to understand the impact of the
    previous AGB evolution for post-AGB phases.}  { We computed a grid
    of post-AGB full evolutionary sequences that include all previous
    evolutionary stages from the zero age main sequence to the white
    dwarf phase. We computed models for initial masses between 0.8 and
    4 $M_\odot$ and for a wide range of initial metallicities
    ($Z_0=$0.02, 0.01, 0.001, 0.0001). This allowed us to provide
    post-AGB timescales and properties for H-burning post-AGB objects
    with masses in the relevant range for the formation of planetary
    nebulae ($\sim 0.5 \mbox{--} 0.8\, M_\odot$).  We included an
    updated treatment of the constitutive microphysics and included an
    updated description of the mixing processes and winds that play a
    key role during the thermal pulses (TP) on the AGB phase.  } { We
    present a new grid of models for post-AGB stars that take into
    account the improvements in the modeling of AGB stars in recent
    decades. These new models are particularly suited to be inputs in
    studies of the formation of planetary nebulae and for the
    determination of the properties of CSPNe from their observational
    parameters.  We find post-AGB timescales that are at least
    approximately three to ten times shorter than those of old
    post-AGB stellar evolution models. This is true for the whole mass
    and metallicity range. The new models are also $\sim 0.1 \mbox{--}
    0.3$ dex brighter than the previous models with similar remnant
    masses. Post-AGB timescales only show a mild dependence on
    metallicity.}  {The shorter post-AGB timescales derived in the
    present work are in agreement with recent semiempirical
    determinations of the post-AGB timescales from the CSPNe in the
    Galactic bulge. The lower number of post-AGB and CSPNe predicted
    by the new models might help to alleviate some of the
    discrepancies found in the literature. As a result of the very
    different post-AGB crossing times, initial final mass relation and
    luminosities of the present models, the new models will have a
    significant impact on the predictions for the formation of
    planetary nebulae and the planetary nebulae luminosity
    function. In particular, the new models should help to understand
    the formation of low-mass CSPNe as inferred from asteroseismic and
    spectroscopic determinations.}


   \maketitle
%

\section{Introduction}

The transition between the asymptotic giant branch (AGB) and white
dwarf phases  is arguably one of the least understood phases of
the evolution of low- and intermediate-mass single stars ($M_{\rm i}
\sim 0.8\mbox{--} 8 M_\odot$). During this phase, stars are expected
to evolve as OH/IR stars, protoplanetary nebula central stars and,
under the right conditions,  the central stars of planetary nebulae 
\citep{2003ARA&A..41..391V, 2005ARA&A..43..435H,2014RMxAA..50..203K}.
While there are many grids of stellar models covering all phases from
the zero age main sequences (ZAMS) to the AGB regime
(e.g., \citealt{2004ApJ...612..168P,2007PASA...24..103K,2007A&A...476..893S,
  2008ApJS..178...89D,2010MNRAS.403.1413K,2009ApJ...696..797C,2011ApJS..197...17C}),
this is not the case for the post-AGB and CSPNe phases of stellar
evolution.  Only two main grids covering the relevant range of final
masses ($M_{\rm f}\sim 0.5\mbox{--}0.8 M_\odot$) are available. Those
are the grids computed by \cite{1994ApJS...92..125V} from the AGB
models of \cite{1993ApJ...413..641V} and the grids computed by
\cite{1995A&A...299..755B} from the AGB models of
\cite{1995A&A...297..727B}, which are usually complemented at low
masses by the models of \cite{1983ApJ...272..708S}.

Other than the fact that they are very interesting objects in
themselves, post-AGB stars are also useful for other fields of
astrophysics \citep{2014RMxAA..50..203K}. In particular, planetary
nebulae (PNe) are bright, easy to identify, and their progenitors are
expected to have ages spanning from $\sim 0.1$ Gyr to $\sim 10$ Gyr.
In the most simple scenario, PNe are formed when the progenitor stars
lose their external layers at the end of the AGB and cross the HR
diagram on their way to the white dwarf cooling sequence. While
crossing the HR diagram the central stars of the PNe (CSPNe) become
sufficiently hot to ionize the previously ejected material
\citep{1957IAUS....3...83S,1966PASP...78..232A,1970AcA....20...47P}. 
Planetary nebulae and CSPNe offer unique insight into the
nucleosynthesis during the TP-AGB phase. Extragalactic PNe can be used
to understand metallicity gradients and their temporal evolution in
galaxies. In addition masses and numbers of extragalactic PNe can be
used to derive stellar formation rates. Last but not least, the PN
luminosity function (PNLF) has proven to be a good distance indicator
as far as $\sim 20$ Mpc, although we still do not understand why
\citep{2012Ap&SS.341..151C}. The formation and detectability of PNe
depends strongly on the relationship between two different timescales:
Evolutionary timescales of the CSPNe, which provide ionizing photons,
and dynamical timescales of the circumstellar material ejected at the
end of the AGB \citep{2001A&A...378..958M}. If the CSPN evolves too
fast the PN is ionized for a short time, and thus has a low detection
probability or might even not be ionized at all. Conversely, if the
star evolves too slowly, the ionization of the nebula takes place when
the ejected material has already dispersed too much to be
detectable. In this work, we address the first of these
timescales. Namely, we present full stellar evolution computations of
the post-AGB and CSPNe phases.  The evolutionary timescale of the
post-AGB remnant is mainly set by the speed at which the H-rich
envelope is consumed before reaching its final value on the WD
stage. Models departing from the TP-AGB with less massive H-rich
envelopes, higher luminosities, or more intense winds must evolve
faster than those with more massive envelopes, lower luminosities, and
less intense winds. In its turn, the post-AGB luminosity and the mass
of the envelope at the departure from the AGB depends on the details
of the TP-AGB evolution. Consequently, proper modeling of the previous
evolutionary phases is needed to obtain accurate post-AGB timescales.

There are some indications that available models of the post-AGB and
CSPN phases are not accurate enough. First, the two available grids of
post-AGB models \citep{1994ApJS...92..125V,1995A&A...299..755B} do not
agree with each other on the predicted timescales
\citep{2008ApJ...681.1296Z}. Second, consistency between the masses of
white dwarfs and those of CSPNe seems to require faster evolutionary
speeds than predicted by both sets of models
\citep{2014A&A...566A..48G}. Third, present models of the CSPNe phase
are unable to explain why the cut-off of the PNe luminosity function
is constant in most galaxies
\citep{2001A&A...378..958M,2004A&A...423..995M}. Lastly,
post-AGB stellar evolution models,  computed with updated physics  in
a reduced mass range \citep{2008PhDT.......290K,2009A&A...508.1343W},
show a strong disagreement with the previous grids. This is not a
surprise since many improvements have been carried out in the field of
stellar physics in recent decades. Most importantly, available
grids have been computed with opacities, which are now 45 years old
\citep{1970ApJS...19..261C,1970ApJS...19..243C} before the big
changes introduced by the OPAL \citep{1996ApJ...464..943I}, and
Opacity Project \citep{2007MNRAS.382..245S}
redeterminations. Similarly, nuclear reaction rates, equation of
states, conductive opacities, and neutrino emission rates adopted in
the models date from the early eighties and even earlier.  In
addition, \cite{1997A&A...324L..81H} showed that the existence of
carbon stars at low luminosities can be explained by the addition of
 mixing beyond the formal convective boundaries during the
thermal pulses (TP) on the AGB.  Finally,
\cite{2002A&A...387..507M} showed that C-rich molecular opacities are
essential to predict the correct effective temperatures once the AGB models
become carbon rich ($N_{\rm C}/N_{\rm O} > 1$, by number
fractions). This is particularly important because of the impact of
effective temperatures on the mass loss rates.  While all these
improvements in stellar modeling have been implemented in AGB stellar
models, and very detailed and exhaustive grids and models are available
\citep{2009A&A...508.1343W,2009ApJ...696..797C,2010MNRAS.408.2476V,
  2010MNRAS.403.1413K,2011ApJS..197...17C,2012ApJ...747....2L,
  2014ApJ...784...56C,2015MNRAS.446.2599D}, the inclusion of these
improvements in post-AGB stellar models is still missing. It is time for a
recomputation of the post-AGB models in the light of all these
advances.

 The  aim of our study is to assess the post-AGB 
timescales and properties with the help of post-AGB stellar evolution
models, which  include an updated treatment of the relevant physics, and
in particular of the AGB phase. In  Sect. 2, we describe
the input physics and calibration of free parameters in the stellar
evolution models. In section \ref{sec:ZAMS-AGB} we show the
agreement of our models with several observables and with the other  
  state-of-the-art AGB models. This shows the reliability of the new
models. In section \ref{sec:post-AGB} we describe the results of
our computations and discuss differences with previous post-AGB
models. Then, in section \ref{sec:discussion} we discuss possible
consequences of our results and the uncertainties behind the present
computations. Finally, we close the article with a summary and some
conclusions.

\section{Input physics, numerics, and set up}
\label{sec:input}
The calculations reported here have been carried out with the last version of {\tt
  LPCODE} stellar evolutionary code. This code has been used to study
different problems related to the formation and evolution of white dwarfs
\citep{2013A&A...557A..19A,2013ApJ...775L..22M,2013A&A...555A..96S,2015A&A...576A...9A}. {\tt
  The LPCODE} is a well-tested stellar evolution code and has been recently tested
against other stellar evolution codes during the main sequence, red giant, and
white dwarf phases \citep{2013A&A...555A..96S}\footnote{Testing of the code at
  the main sequence and red giant phases was performed in a series of
  workshops ``The Aarhus Red Giants Workshops'', which can be found at {\tt
    http://users-phys.au.dk/victor/rgwork/}}. The numerical methods adopted in
{\tt LPCODE} are extensively described in
\cite{2003A&A...404..593A,2005A&A...435..631A}. Recent improvements in the
numerical scheme, as well as convergence problems and how we circumvented
them, are briefly described in Appendix \ref{app:numerics}. In what follows,
we describe the adopted micro- and macrophysics in the present work.

\subsection{Microphysics}
  We have adopted  state-of-the-art
ingredients for the microphysics relevant for the evolution and structure of
low- and intermediate-mass stars to supersede previous post-AGB grids. The nuclear network accounts explicitly for
the following elements: $^{1}$H, $^{2}$H, $^{3}$He, $^{4}$He, $^{7}$Li,
$^{7}$Be, $^{12}$C, $^{13}$C, $^{14}$N, $^{15}$N, $^{16}$O, $^{17}$O,
$^{18}$O, $^{19}$F, $^{20}$Ne, and $^{22}$Ne, together with 34 thermonuclear
reaction rates for the pp-chains, CNO bi-cycle, and helium burning. These reaction rates are
identical to those described in \cite{2005A&A...435..631A} with the exception
of the reactions $^{12}$C$\ +\ $p$ \rightarrow \ ^{13}$N + $\gamma \rightarrow
\ ^{13}$C + e$^+ + \nu_{\rm e}$ and $^{13}$C(p,$\gamma)^{14}$N, which are
taken from \cite{1999NuPhA.656....3A}, and the reaction rate
$^{14}$N(p,$\gamma)^{15}$O, which was taken from \cite{2005EPJA...25..455I}.
High temperature radiative opacities are taken from
\citep{1996ApJ...464..943I} and conductive opacities are taken from OPAL from
\cite{2007ApJ...661.1094C}, respectively. We use updated low-temperature molecular
opacities with varying C/O ratios. For this purpose, we have adopted
the low-temperature opacities of \cite{2005ApJ...623..585F}  extended with the tables for varying C/O ratios  presented  in
\cite{2008PhDT.......290K} and \cite{2009A&A...508.1343W}. In {\tt LPCODE,}
molecular opacities are computed by adopting the opacity tables with the
correct abundances of the unenhanced metals (e.g., Fe) and C/O ratio.
Interpolation is carried out by means of separate quadratic interpolations in
$R=\rho/{T_6}^3$, $T$ and $X_{\rm H}$, but linearly in $N_{\rm C}/N_{\rm O}$.
This approach is inferior to the on the fly computation of opacities
\citep{2013MNRAS.434..488M}, but allows us to capture the first order effect
of the formation of a C-rich envelope without the need to follow the huge
number of ions and elements that participate in the opacity of AGB
envelopes. The equation of state during the main-sequence evolution is the
updated {\tt EOS\_2005}\footnote{Available at {\tt
    http://opalopacity.llnl.gov/EOS\_2005/}} version of the {\tt OPAL EOS}
\citep{1996ApJ...456..902R} for H- and He-rich composition and a given
metallicity. Neutrino emission rates for pair, photo, and bremsstrahlung
processes are those of \cite{1996ApJS..102..411I}, while plasma processes are
included with the expressions presented by \cite{1994ApJ...425..222H}. For the
early white dwarf regime, we use the equation of state of
\cite{1979A&A....72..134M} for the low-density regime, while for the
high-density regime we consider the equation of state of
\cite{1994ApJ...434..641S}. Outer boundary conditions are set by simple
Eddington gray $T(\tau)$-relations.

\subsection{Macrophysics}
\label{sec:Macrophysics}
While the microphysics of stellar models is relatively well established, 
macrophysical processes are their main uncertainty. In particular, the modeling
of convective and nonconvective mixing processes and stellar winds are among
the main uncertainties in the computation of stellar evolution sequences. This
is even worst in the case of stellar evolution computations from the ZAMS all the way down to the white dwarf stage, such as those
presented in this work. In these computations one must deal with many
convective regions and with winds during many, and very different, stages of
the evolution; see \cite{2013EPJWC..4301002W} for a nice review on this
topic. In what follows, we describe the prescriptions adopted for the stellar
winds during the different stages of the evolution and how we have calibrated
the different free parameters  involved in the treatment of convection.

\subsubsection{Stellar winds}

Mass loss during the RGB is usually included in stellar evolutionary
computations following the empirical formula of
\cite{1975MSRSL...8..369R}. More recently,
\citep{2005ApJ...630L..73S,2007A&A...465..593S} argued for a
reinterpretation of this formula in terms of a more physical picture of the
mechanism behind mass loss. This  led to the update of Reimers' formula
including two more factors.  \cite{2010ApJ...724.1030G} and
\cite{2014ApJ...790...22R} showed that this new formula also provides a
better description of predust AGB winds. In line with these recent studies, we
have included winds from cold giants following the prescription of
\cite{2005ApJ...630L..73S}, i.e.,
\begin{equation}
\frac{\dot{M}^{\rm SC}}{M_\odot/{\rm yr}}=8\times10^{-14}\frac{\left(L_\star/L_\odot\right)\left(R_\star/R_\odot\right)}
{\left(M_\star/M_\odot\right)} \left(\frac{T_{\rm eff}}{\rm 4000  K}\right)^{3.5}
\left(1+\frac{g_\odot}{4300  g_\star}\right).
\label{eq:sk}
\end{equation}
Since the seminal work of \cite{1979A&A....79..108S}, it became
evident that steady stellar winds play a decisive role in the AGB. In
particular, it is stellar winds that rule the length of the thermally
pulsating AGB phase (TP-AGB). Stellar winds during the AGB lead to the
(almost) complete removal of the H-rich envelope, forcing the remnant
star to contract to the white dwarf phase. Yet, AGB stellar winds are
not fully understood. While theoretical and observational evidence of
the existence of pulsation-enhanced, dust-driven winds is strong for
C-rich stars, the situation of O-rich (M-type) AGB stars is much less
clear. In particular, numerical simulations of pulsation-enhanced,
dust-driven winds are unable to find efficient mass loss rate
\citep{2006A&A...460L...9W}, although there is still hope for this
mechanism
\citep{2012Natur.484..220N,2012Natur.484..172H,2015A&A...575A.105B}.

The inclusion of C-rich molecular opacities implies a different
treatment of O- and C-rich AGB stars, as both radius ($R_\star$) and
effective temperature ($T_{\rm eff}$) are very sensitive to the C/O
ratio of the model. In its turn, pulsation-enhanced, dust-driven winds
are themselves very sensitive to the values of $R_\star$ and $T_{\rm
  eff}$. Then, a consistent treatment of pulsation-enhanced
dust-driven winds with the actual C/O ratio is needed.  In order to
have a internally consistent description of AGB winds we  cannot
rely on theoretical determinations, as those are unavailable for
O-rich stars. Fortunately,
\cite{1998MNRAS.293...18G,2009A&A...506.1277G} has determined mass
loss rates for both C- and M- type AGB stars adopting the same
techniques. Following the suggestions by
\cite{1998MNRAS.293...18G,2009A&A...506.1277G}, we adopted for
pulsating O-rich AGB stars the relation
\begin{equation}
\log{\frac{\dot{M}^{\rm O}}{M_\odot/{\rm yr}}}=-9+0.0032\, (P/{\rm day}),
\label{eq:Mdot-MAGB}
\end{equation}
while the mass loss for the winds of pulsating carbon stars was adopted as
\begin{equation}
\log{\frac{\dot{M}^{\rm C}}{M_\odot/{\rm yr}}}=-16.54+4.08\, \log(P/{\rm day}).
\label{eq:Mdot-CAGB}
\end{equation}
To compute the mass loss from eqs. \ref{eq:Mdot-MAGB} and
\ref{eq:Mdot-CAGB}, it is necessary to estimate the value of the pulsation
period $P$. We compute
$P$ from the relation  of \cite{1986ApJ...311..864O}, i.e.,
\begin{equation}
\log(P/{\rm day})=-1.92-0.73\log(M_\star/M_\odot)+1.86\log(R_\star/R_\odot),
\label{eq:PulPeriod}
\end{equation}
which for values of $\log(R_\star/R_\odot)\gtrsim 2.5$ and
$1<M_\star/M_\odot<10$ is always within a 10\% of the $P(M,R)$ relation of
\cite{1990fmpn.coll...67W} (as given by \citealt{1993ApJ...413..641V}).

It is well known that there must be some kind of upper limit for the intensity
of the winds. It has  usually been argued that pulsation-enhanced,
dust-driven winds must be constrained by the single scattering limit
(e.g., \citealt{1994ApJS...92..125V}), i.e., the situation in which all the
momentum of the stellar radiation field is transferred to the wind
\begin{equation}
\dot{M}^{\rm SS\, lim}=\frac{L_\star}{c\, v_\infty},
\label{eq:SSLim}
\end{equation}
 where $c$ is the speed of light and $v_\infty$ is the terminal wind
 velocity, for which we assume $v_\infty= 10$km/s as suggested by the
 results of \cite{2009A&A...506.1277G}.  However, as argued by
 \cite{2010A&A...509A..14M}, there are theoretical reasons to think
 that such a limit is not appropriate for pulsation-enhanced,
 dust-driven winds.  In fact, the numerical simulations from
 \cite{2010A&A...509A..14M} show many models with mass loss rates
 beyond the single scattering limit given by eq. \ref{eq:SSLim}. In
 addition, more recent observational evidence
 \citep{2013lcdu.confE..94G} supports the existence of mass loss rates
 beyond the single scattering limit. Following the results presented
 in \cite{2010A&A...509A..14M} we constrained the mass loss
 rates of eqs. \ref{eq:Mdot-MAGB} and \ref{eq:Mdot-CAGB} not to exceed
 three times the value given by eq. \ref{eq:SSLim}. This is a rather
 arbitrary choice, but in line with the results presented in
 \cite{2010A&A...509A..14M} and \cite{2013lcdu.confE..94G}.

Finally, during the CSPN phase, radiation-driven winds must also be
included. Following \cite{1995A&A...299..755B}, we derived a similar
relation from the results of \cite{2004A&A...419.1111P}, i.e.,
\begin{equation}
\frac{\dot{M}^{\rm CSPN}}{M_\odot/{\rm yr}}=9.778\times10^{-15}\times
\left(L_\star/L_\odot\right)^{1.674}\times\left(Z_0/Z_\odot\right)^{2/3},
\label{eq:cspn}
\end{equation}
which is, for the range of luminosities of interest, always within a factor of
two from the mass loss rates derived by \cite{1995A&A...299..755B} from the
earlier results of \cite{1988A&A...207..123P}.  Given that our grid spans a
factor 20 in the initial metal content of the stars, the dependence with $Z_0$
in eq. \ref{eq:cspn} was included after the derivation. The factor
${Z_0}^{2/3}$ reproduces the known dependence of radiation-driven winds with
the heavy metal content of the star, i.e., mostly of iron; see
\citep{2001A&A...369..574V}.

Finally, all these prescriptions (eqs. \ref{eq:sk}, \ref{eq:Mdot-MAGB},
\ref{eq:Mdot-CAGB}, \ref{eq:SSLim} and \ref{eq:cspn}) for stellar winds in
different regimes must be combined. This is done under the assumption that
the mechanism driving cool winds in RGB stars is always active until
pulsation-driven winds develop. Specifically, once $P>100$ days, we take the
cool wind rate to be the maximum between eqs \ref{eq:sk} and \ref{eq:Mdot-MAGB}
(\ref{eq:Mdot-CAGB}), when $N_{\rm C}/N_{\rm O}<1$ ($N_{\rm C}/N_{\rm O}>1$)
\begin{eqnarray}
\dot{M}^{\rm cool}&=& {\rm Max}(\dot{M}^{SC},\dot{M}^{\rm AGB}), {\rm if\  P>100 d} \\
`&=& \dot{M}^{SC}, {\rm if\ P<100 d}.\\
\end{eqnarray}
Where $\dot{M}^{\rm AGB}$ stands for $\dot{M}^{\rm C}$ or $\dot{M}^{\rm
  O}$, depending on whether $N_{\rm C}/N_{\rm O}$ is above or below unity, 
 and constrained not to exceed three times the single scattering limit.

While all mass loss prescriptions are somewhat uncertain, the mass
loss rates during the transition from the cold AGB to the hot CSPN
phase is completely unconstrained. \cite{2007ASPC..378..343S} argue
that cool winds must last until $T_{\rm eff}\sim 5000\mbox{--} 6000$K to
reproduce the spectral energy distribution of post-AGB objects.  We
have chosen to exponentially decrease the cool wind rates to its CSPN
values  (eq. \ref{eq:cspn}), as the model evolves from $\log T_{\rm
  eff}=3.8$ to $\log T_{\rm eff}=4.1$, by means of a simple linear
interpolation in the logarithmic rates
\begin{equation}
\log \dot{M}^{\rm trans}= x\,\log \dot{M}^{\rm  cool}+(1-x)\,\log\dot{M}^{\rm CSPN},
\label{eq:interpol}
\end{equation}
where $x=(\log T_{\rm eff}-3.8)/0.3$.  The accuracy of
this interpolation between the extrapolation of two prescriptions
outside their validity range is completely questionable. However, we
show in the following sections that, unless mass loss rates are
much higher than these values, they are of no relevance for post-AGB evolution.

\subsubsection{Convection and convective boundary mixing}
We treat convection according to the \cite{2012sse..book.....K}
formalism of mixing length theory (MLT;
\citealt{1932ZA......5..117B,1958ZA.....46..108B}). The MLT free
parameter $\alpha_{\rm MLT}$ has been fixed with solar calibration and
kept constant for all masses during all evolutionary phases.  For the
solar calibration, one needs to find  the combination of initial
He (Y) and metal (Z) mass fractions and the value of the mixing-length
parameter that reproduce the solar radius ($R_\odot=6.96\times
10^{10}$cm), luminosity ($L_\odot=3.842\times 10^{33}$ erg s$^{-1}$;
\citealt{1995RvMP...67..781B}), and ratio $Z/X=0.0245\pm 0.005$
\citep{1993oee..conf...15G} at an age of $t_\odot= 4.57$ Gyr. Solar
models without microscopic diffusion or with new
\cite{2009ARA&A..47..481A} compositions cannot properly account for
some seismic properties of the Sun. For this reason, we decided to
account for atomic diffusion and to adopt the
\cite{1993oee..conf...15G} initial chemical composition for the
calibration of the sun. Once the calibration of the mixing length and
initial composition is carried out, the accuracy of the solar model
can be tested by comparing the depth of the outer convective layers
and surface $Y$ value with the values obtained from helioseismological
studies ($Y_\odot^{surf}=0.2485\pm 0.0035$; $R_\odot^{CZ}=0.713\pm
0.001 R_\odot$; \citealt{2009ApJ...705L.123S}).  Our treatment of time
dependent element diffusion is based on the multicomponent gas picture
of \cite{1969fecg.book.....B} taking the effects of gravitational
settling, chemical diffusion, and thermal diffusion into account, but
neglecting radiative levitation. In particular, we solved the
diffusion equations within the numerical scheme described in
\cite{2003A&A...404..593A}.

We computed several $1 M_\odot$ models starting from the ZAMS and
adopting different values of the initial composition and mixing length. The
best solar model was obtained by assuming $X_{\rm ini}=0.7092$, $Z_0=0.0194$ and a value of the MLT parameter of $\alpha_{\rm
  MLT}=1.825$. With these values, we obtain values of $L=0.9994 L_\odot$,
$T_{\rm eff}=5776.85$K, $Y^{surf}=0.24415$, $(Z/X)^{surf}=0.02488,$ and
$R^{CZ}=0.7144 R_\odot$ at an age of $t=4.5684$ Gyr. We consider this an
overall good solar model and, consequently, we adopt $\alpha_{\rm MLT}=1.825$
throughout this work.

Turbulent mixing beyond the formal convective boundaries described by
a bare Schwarzschild criterion \citep{1906WisGo.195...41S} is one of
the main uncertainties in stellar astrophysics. From now on, we
  call this process convective boundary mixing (CBM). In {\tt LPCODE,}
 CBM processes are included following the suggestion of
\cite{1996A&A...313..497F} of an exponentially decaying velocity
field. The diffusion coefficient beyond convective boundaries is
\citep{1997A&A...324L..81H}
\begin{equation}
D_{\rm OV }=D_0\times \exp\left[\frac{-2z}{f H_P}\right],
\label{eq:OV}
\end{equation}
where $D_0$ is the diffusion coefficient provided by the MLT
($D_0=v_{\rm MLT}\alpha_{\rm MLT}H_P$) close to the convective
boundary, $H_P$ is the pressure scale height at the convective
boundary, $z$ is the geometric distance to the formal convective
  boundary, and $f$ is a free parameter that must be calibrated.  We
take $D_0$ as the mean value of the MLT diffusion coefficient in the
region within $0.1 H_P$ from the  formal convective boundary and
the  CBM region is extended until the diffusion coefficient falls
10 orders of magnitude, i.e., $D_{\rm cut-off}=10^{-10}D_0$. The value $D_{\rm cut-off}$ (which is rarely stated
in the literature) is as important as the value of $f$ in slow
evolutionary stages, where mixing of the main chemical elements is
always complete and only the extension of the  CBM region,
and not the mixing speed, becomes relevant (e.g., core H- and
He-burning).

For our aims, the main convective boundaries (and
related $f$ values) are the boundaries of convective cores during core H-
and He-burning ($f_{\rm CHB}$ and $f_{\rm CHeB}$),  the lower 
boundaries of the convective envelope ($f_{\rm CE}$), and the pulse-driven
convective zone ($f_{\rm PDCZ}$) during the thermal pulses on the AGB.

There is a general agreement, within uncertainties, that an extension of
$0.2\, H_P$ in the convective core of upper main-sequence stars offers a
relatively good agreement with observation of the main sequence in open
clusters and the field
\citep{1991A&AS...89..451M,1992ApJ...390..136S,1992A&AS...96..269S,1997A&A...324L..81H,2004ApJ...612..168P,2009A&A...508.1343W,2012A&A...537A.146E}. With
our choice of $D_{\rm cut-off}$, this is reproduced by assuming a value of
$f_{\rm CHB}\sim 0.0174$.  This value of the extension of the convective
core is in good agreement with determinations coming from eclipsing binaries; see \cite{2007A&A...475.1019C,2015A&A...575A.117S}.  It is well known that
the inclusion of  CBM processes in small convective cores of low-mass
main-sequence stars must be restricted. Following previous works
\citep{2004ApJ...612..168P,2009A&A...508.1343W}, we  adopted a linear
dependence of $f_{\rm CHB}$ with the initial mass of the star, so that it
attains $f_{\rm CHB}= 0.0174$ in the upper main sequence ($M>M_2$) and
decreases to zero for stars without convective cores ($M<M_1$) (see table
\ref{tab:fZAMS}). Following \cite{2012A&A...537A.146E}, we checked that
this choice allows us to reproduce the width of the upper main sequence as
presented by \cite{1997PASP..109..759W}.

\begin{table}
  \caption{$M_1$ and $M_2$ values for the boundary mixing recipe at the
    convective core in the main sequence, i.e., $f_{\rm
      CHB}=0.0174\times(M_i-M_1)/(M_2-M_1)$ for $M_i$ between $M_1$ and $M_2$}          
\label{tab:fZAMS}     
\centering         
\begin{tabular}{c c c}     \hline\hline                
$Z_0$ & $M_1/M_\odot$ & $M_2/M_\odot$ \\   \hline                       
0.02 &  1.15    & 1.75 \\
0.01 &  1.15    & 1.75 \\
0.001 &  1.3    & 1.75 \\
0.0001 &  1.6    & 2.2 \\ \hline                                  
\end{tabular}
\end{table}

The  CBM at the border of the He-burning core is significantly less
understood. While there are some physical arguments
(e.g., \citealt{1985ApJ...296..204C}) and asteroseismic inferences
\citep{2013EPJWC..4304005C, 2015arXiv150601209C} in favor of 
 the existence of CBM at convective boundaries, its extension is much less
constrained. In the absence of a better constraint, we  set $f_{\rm CHeB}=
0.0174$ (i.e., $= 0.2\, H_P$) in our simulations for all stars.

Of particular interest for the present work are the  CBM
processes during the thermal pulses on the AGB. In {\tt LPCODE} and
in many other codes
(e.g., \citealt{1997A&A...324L..81H,2009A&A...508.1343W} and references
therein) third dredge up (3DUP) does not appear, or is not very
efficient, in low-mass (luminosity) AGB models without  CBM
processes. This contradicts the existence of  carbon stars
($N_{\rm C}/N_{\rm O}>1$ by number fractions) in the lower AGB and at all
metallicities. Furthermore,  CBM processes at the lower boundary
of the convective envelope ($f_{\rm CE}$) are also needed to create
the $^{13}$C pockets needed for the creation of s-process
elements. Also,  CBM at the lower boundary of the pulse-driven
convective zone ($f_{\rm PDCZ}$) is needed to reproduce the oxygen
abundances of post-AGB PG1159 stars, which are believed to display the
intershell abundances of AGB stars \citep{1999A&A...349L...5H}.  The
values of $f_{\rm CE}$ and $f_{\rm PDCZ}$ are certainly not well
constrained. The early exploration of \cite{2000A&A...360..952H}
suggested that $0.01\lesssim f_{\rm PDCZ}\lesssim 0.03$ would
reproduce the abundances of PG1159 stars, while $f_{\rm CE}$ should be
significantly larger. Studying the production of s-process elements,
\cite{2003ApJ...586.1305L} suggested that these values should be of
$f_{\rm PDCZ}\sim 0.008$ and $f_{\rm CE}\sim 0.128$.
\cite{2005ARA&A..43..435H} argued that the oxygen abundances in PG1159
stars can be reproduced with values of $0.005\lesssim f_{\rm
  PDCZ}\lesssim 0.015$ and that a value of $f_{\rm CE}\sim 0.13$ would
be enough to account for the s-process production, but would lead to
neutron exposures that are too large and an overabundance of second peak
elements. Unfortunately, it has been shown by
\cite{2009ApJ...692.1013S} and \cite{2009A&A...508.1343W} that the
inclusion of  CBM processes both at the convective envelope and
the pulse-driven convective zone leads to an initial-final mass
relation (IFMR) in disagreement with observations. This is
particularly true for initial masses of $\sim 3M_\odot$, which would have final masses of $\sim 0.6 M_\odot$. These masses are far from the final masses of $\gtrsim 0.7
M_\odot$ suggested by semiempirical determinations of the IFMR from
stellar clusters (\citealt{2008MNRAS.387.1693C},
\citealt{2009ApJ...692.1013S}, \citealt{2014A&A...566A..48G} and
references therein).

We performed several exploratory computations with {\tt LPCODE} that
confirm the previous picture. We find that values of $0.005\lesssim
f_{\rm PDCZ}\lesssim 0.01$ are needed to reproduce the high O
abundances of PG1159 stars, while the simultaneous inclusion of
$f_{\rm CE}\gtrsim 0.002$ leads to final masses not in agreement with
our expectation from semiempirical IFMRs.  In fact, in a preliminary
study (\citealt{2015ASPC..493...83M}; from now on M15, see Appendix
\ref{app:M15}), we adopted values of $f_{\rm PDCZ}= 0.005$ and $f_{\rm
  CE}= 0.13,$ and the resulting IFMR of the theoretical models is far
from that suggested by Galactic clusters at solar metallicity, and
even stars with $M\lesssim 1 M_\odot$ become carbon stars, at variance
with observations of stellar clusters in the Magellanic Clouds
\citep{2007A&A...462..237G}. In the next section, we show that
choosing $f_{\rm PDCZ}= 0.0075$ and $f_{\rm CE}= 0$ allows us to
reproduce many relevant AGB and post-AGB observables.  Note that we do
not claim that $f_{\rm CE}$ should be $f_{\rm CE}= 0$ at all
times\footnote{The existence of the $^{13}$C pocket needed for
  s-process nucleosynthesis implies the need for a layer with both H
  and $^{12}$C after the 3DUP. While the problem with the IFMR comes
  from the inclusion of CBM at the bottom of the convective envelope
  before and during the development of the 3DUP. As the entropy
  barrier drops during the thermal pulse, it seems plausible that CBM
  or other mixing processes are not equally effective before and at
  the end of the thermal pulse.}, but only that this choice of
parameters allow us to reproduce many relevant AGB and post-AGB
observables. Among them are the C/O ratios of AGB and post-AGB stars
in the Galactic disk, the IFMR at near solar metallicities, the C/O
intershell abundances of AGB stars as observed in PG1159 stars, and
the mass range of C-rich stars in the stellar clusters of the
Magellanic Clouds. In addition, the resulting AGB stellar models show
properties (core growth, carbon enrichment, and IFMR) well within the
predictions of available AGB grids. Then, the predicted structures and
timescales obtained with this choice of $f_{\rm PDCZ}$ and $f_{\rm
  CE}$ can be considered good representatives of state-of-the-art AGB
stellar evolution modeling for the post-AGB phase.

\section{Evolution from the ZAMS to the end of the AGB}
\label{sec:ZAMS-AGB}

\begin{table*}
  \caption{Main properties of the sequences from the ZAMS to 
    TP-AGB.}
\label{tab:ZAMS-TPAGB}
\centering
\begin{tabular}{cccccccccccc} 
  \hline\hline             
  $M_i$ & $\tau_{MS}$ & $\tau_{RGB}$ & HeCF & $\tau_{HeCB}$  &  $\tau_{eAGB}$ & $M_c^{1TP}$ &  $\tau_{TP-AGB(M)}$& $\tau_{TP-AGB(C)}$& \#TP & $M_f$ & $N_{\rm C}/N_{\rm O}$\\
  $[M_\odot]$ & [Myr] &[Myr]   &  & [Myr]       &  [Myr]     &$[M_\odot]$& [Myr]             & [Myr]&     (AGB)       &      $[M_\odot]$  & \\
  \hline
  \multicolumn{12}{c}{$Z_0=0.02$}\\ 
  \hline
  1.00     &   9626.5     &   2043.9  &  yes     &   131.60     &   11.161     &   0.5119     &  0.65512     &   0.0000     &  4 &   0.5281     & .400     \\
  1.25     &   3857.6     &   1262.8  &  yes     &   124.96     &   10.302     &   0.5268     &  0.89729     &   0.0000     &  6 &   0.5615     & .368     \\
  1.50     &   2212.4     &   480.67  &  yes     &   114.64     &   11.508     &   0.5267     &   1.2959     &  0.16583E-01 & 11 &   0.5760     & 2.20     \\
  2.00     &   1055.0     &   33.928  &  no     &   307.42     &   19.420     &   0.4873     &   2.9196     &  0.97759E-01 & 18 &   0.5804     & 1.12     \\
  3.00     &   347.06     &   4.6762  &  no     &   83.942     &   6.2381     &   0.6103     &  0.79992     &  0.10622     & 18 &   0.6573     & 1.66     \\
  4.00     &   163.85     &   1.5595  &  no    &   28.196     &   1.9150     &   0.8086     &  0.32029     &   0.0000     & 34 &   0.8328     & .812 \\
  \hline
  \multicolumn{12}{c}{$Z_0=0.01$}\\ 
  \hline
  1.00     &   7908.5     &   1697.2  &  yes     &   113.53     &   11.965     &   0.5128     &  0.73881     &   0.0000     &  4 &   0.5319     & .372     \\
  1.25     &   3259.5     &   988.43  &  yes     &   110.20     &   11.177     &   0.5276     &   1.0972     &  0.14698E-01 &  8 &   0.5660     & 3.93     \\
  1.50     &   1889.2     &   400.18  &  yes     &   101.54     &   11.439     &   0.5289     &   1.2904     &  0.75163E-01 & 10 &   0.5832     & 1.13     \\
  2.00     &   921.34     &   27.072  &  no     &   282.61     &   12.747     &   0.5170     &   1.6179     &  0.37651     & 13 &   0.5826     & 1.98     \\
  2.50     &   509.27     &   8.6930  &  no     &   148.01     &   7.6400     &   0.5728     &  0.71424     &  0.41896     & 12 &   0.6160     & 2.48     \\
  3.00     &   318.67     &   4.0683  &  no     &   73.307     &   4.1321     &   0.6755     &  0.31492     &  0.14332     & 13 &   0.7061     & 1.97     \\
  \hline
  \multicolumn{12}{c}{$Z_0=0.001$}\\ 
  \hline
  0.800     &   12430.     &   1313.5  &  yes     &   116.80     &   13.089     &   0.4918     &  0.35321     &   0.0000     &  0 &   0.4971     & .271     \\
  0.900     &   8020.9     &   1024.7  &  yes     &   102.38     &   10.345     &   0.5167     &  0.77626     &   0.0000     &  3 &   0.5340     & .316     \\
  1.00     &   5413.9     &   808.66  &  yes     &   95.343     &   10.531     &   0.5282     &  0.85648     &  0.75481E-01 &  4 &   0.5517     & 6.61     \\
  1.25     &   2368.4     &   533.69  &  yes     &   103.72     &   8.2357     &  0.5439     &   1.0575     &  0.15402     &  7 &   0.5849     & 5.89  \\
  1.50     &   1300.9     &   288.15  &  yes     &   99.209     &   7.7074     &   0.5573     &  0.74257     &  0.46337     &  8 &   0.5995     & 5.79     \\
  1.75     &   951.65     &   46.973  &  no$^\dagger$  &   210.27     &   8.2381     &   0.5298     &   1.1446     &   1.1904     & 15 &   0.5867     & 6.41     \\
  2.00     &   671.98     &   17.363  &  no     &   151.06     &   7.5069     &   0.5717     &  0.60336     &   1.0910     & 15 &   0.6182     & 6.59     \\
  2.50     &   389.98     &   5.9308  &  no     &   70.605     &   3.6796     &   0.6976     &  0.11696     &  0.49306     & 16 &   0.7101     & 6.98     \\
  3.00     &   255.51     &   2.9146  &  no     &   41.077     &   2.0364     &   0.8244     &  0.22804     &  0.82625E-01 & 24 &   0.8314     & 2.38     \\
  \hline
  \multicolumn{12}{c}{$Z_0=0.0001$}\\ 
  \hline
  0.800     &   11752.     &   891.50  &  yes     &   98.484     &   10.089     &   0.5049     &  0.77058     &   0.0000     &  3 &   0.5183     & .194     \\
  0.850     &   9425.1     &   768.48  &  yes     &   93.850     &   9.5246     &   0.5128     &   1.1515     &  0.28265E-01 &  4 &   0.5328     & 6.45     \\
  1.00     &   5255.3     &   525.95  &  yes     &   94.476     &   8.3594     &   0.5280     &   1.3575     &  0.92408E-01 &  5 &   0.5631     & 7.00     \\
  1.50     &   1261.4     &   241.51  &  yes     &   112.01     &   8.1327     &   0.5526     &  0.76920     &  0.89965     & 12 &   0.6024     & 6.50     \\
  2.20     &   470.74     &   10.813  &  no     &   98.665     &   3.2623     &   0.7022     &  0.12220     &  0.48387     & 16 &   0.7130     & 4.86     \\
  2.50     &   349.85     &   5.8835  &  no     &   66.012     &   2.6371     &   0.7568     &  0.20994     &  0.20654     & 17 &   0.7543     & 5.70     \\
  \hline
  \multicolumn{12}{p\textwidth}{$M_i$: Initial mass of the model (at ZAMS).  $\tau_{MS}$:
    Duration of the main sequence until $X^{\rm center}_{\rm
      H}=10^{-6}$. $\tau_{RGB}$: Lifetime from the end of the main
    sequence to He-ignition, set at $\log L_{\rm He}/L_\odot=1$.
    HeCF: Full He-core flash (and subflashes) at the beginning of the
    core He-burning phase. $\tau_{HeCB}$: Lifetime of core He-burning
    until $X^{\rm center}_{\rm He}=10^{-6}$.  $\tau_{eAGB}$: Lifetime
    of the early AGB phase from the end of core helium burning to the
    first thermal pulse. $M_c^{1TP}$: Mass of the H-free core at the
    first thermal pulse (defined as those regions with $X_{\rm
      H}<10^{-4}$). $\tau_{TP-AGB(M)}$: Lifetime of the star in the
    TP-AGB as an M-type star ($N_{\rm C}/N_{\rm O}<1$). $\tau_{TP-AGB(C)}$: Lifetime of
    the star in the TP-AGB as a carbon star ($N_{\rm C}/N_{\rm O}>1$). \#TP: Number of
    thermal pulses on the AGB. $M_f$: Final mass of the
    star. $N_{\rm C}/N_{\rm O}:$ C/O ratio in number fraction at the end of
    the TP-AGB phase.}\\
  \multicolumn{12}{l}{\small  ``no$^\dagger$'' indicates that
    a mild He-burning runaway appeared  but no full He-core
    flash and subflashes  finally  developed.}
\end{tabular}
\end{table*}

    We computed a grid of 27 sequences with initial masses ($M_i$)
    between 0.8M$_\odot$ and 4M$_\odot$ and four different
    metallicities to study the evolution of stars after the AGB at
    different masses and metallicities.  The final masses ($M_f$) of
    the model sequences are between $\sim 0.5 M_\odot$ and $\sim 0.85
    M_\odot$, correspondingly. This is the main range of interest for
    the formation of PNe. The initial metal content of the star
    ($Z_0$) is taken to be solar scaled with overall metal mass
    fractions of $Z_0=$0.02, 0.01, 0.001 and 0.0001.  We follow the
    suggestion by \cite{2009A&A...508.1343W} for the initial helium
    ($Y$) content and we take it to be
\begin{equation}
Y(Z_0)=0.245+Z_0 \times 2,
\end{equation}
in good agreement with our present understanding of Big Bang
nucleosynthesis
\citep{2007ARNPS..57..463S,2013A&A...558A..57I,2013JCAP...11..017A,2014JCAP...10..050C}
and close to our calibration of the solar model.

   The evolution of low- and intermediate-mass stars up to the beginning of the
   TP-AGB is relatively well understood. An updated description of the main
   relevant phases and the physics involved can be found in
   \cite{2012sse..book.....K}. 
   Table \ref{tab:ZAMS-TPAGB} shows the main characteristics of our
   sequences from the ZAMS to the end of the TP-AGB phase. Lifetimes
   during the main sequence, the red-giant branch, and the He-core
   burning stage are denoted by $\tau_{MS}$, $\tau_{RGB}$, and
   $\tau_{HeCB}$, respectively. As is well known, timescales become
   longer as initial mass decreases and metallicity increases. The
   main exception to this general trend is observed around the
   transition from degenerate to nondegenerate He-ignition as stellar
   mass increases. Column 4 in Table \ref{tab:ZAMS-TPAGB} indicates
   whether He-ignition happened in the form of a He-core flash (HeCF)
   or did not occur for each sequence.  As shown in column 5 of Table
   \ref{tab:ZAMS-TPAGB}, $\tau_{HeCB}$ attains a local maximum around
   $M_i\sim 2M_\odot$ for solar-like metallicities
   ($Z_0=0.02,\ 0.01$), and at slightly lower $M_i$ for lower
   metallicities. Below this maximum, sequences that undergo an HeCF
   have an almost constant He-burning lifetime of $\tau_{HeCB}\sim
   10^8$ yr, irrespective of initial mass and metallicity.  In
   contrast, sequences of higher masses show the typical trend of
   decreasing lifetimes with increasing mass.  The behavior of
   $\tau_{HeCB}$ is due to the different sizes of the H-free core
   (HFC) during He-core burning.  The HFC at the beginning of
   He-burning is approximately constant for sequences that undergo a
   HeCF. The HFC decreases at the transition between degenerate and
   nondegenerate He-ignition and monotonically increases with
   increasing initial mass; see for example
   \cite{2013ApJ...766..118M}.  The mass of the HFC during He-burning
   not only affects $\tau_{HeCB}$, but also the properties of the next
   evolutionary stages. The mass and also the composition of the HFC
   keeps a memory of the previous evolutionary history. For example,
   we see in Table \ref{tab:ZAMS-TPAGB} that the duration from the end
   of He-core burning to the first thermal pulse (early AGB phase;
   $\tau_{eAGB}$) also reflects, to a certain extent, the different
   HFC sizes left by He-core burning.  More importantly, for the same
   reason, the HFC at the first thermal pulse shows a clear minimum as
   a function of $M_i$ (see column 7 in Table
   \ref{tab:ZAMS-TPAGB}). This is why models with very efficient 3DUP
   \citep{2009A&A...508.1343W,2015ASPC..493...83M} predict a very
   pronounced plateau in the IFMR below $M_i\sim 2.5M_\odot$.

   In the last two decades, the TP-AGB has been the object of
   many detailed studies (see
   \citealt{2005ARA&A..43..435H,2007PASA...24..103K,
     2009ApJ...696..797C, 2014PASA...31...30K} and references
   therein). Grids of models and yields for TP-AGB stars are now
   available for a wide range of masses and metallicities; e.g., \cite{2010MNRAS.403.1413K,2011ApJS..197...17C}. A detailed
   discussion of the present models in the TP-AGB phase would take us
   too far afield from the goal of the present work, and add almost
   nothing to the present knowledge in the field. Yet, as we discuss in the next sections and  analyze in Appendix
   \ref{app:physics}, the main results of the present work are related
   to the accurate modeling of the TP-AGB phase. Specifically, the
   post-AGB sequences are sensitive to the properties of the HFC and
   the CNO content of the envelope.  Both the properties of the core
   and the envelope of post-AGB stars are determined by 3DUP episodes; see \citealt{2005ARA&A..43..435H} for a
   detailed description of the process.  Third dredge-up intensity not
   only determines the amount of  carbon pollution of the
   envelope but also the IFMR, especially at $M_i\lesssim 3 M_\odot$
   \citep{2009ApJ...692.1013S}. The same post-AGB mass $M_f$ is
   attained by models with very different initial masses and
   evolutionary histories depending on the intensity of 3DUP.  It is
   then important to know to which extent the present models offer an
   accurate description of the main structural properties of AGB
   stars. In the following paragraphs we show that with the
   calibration of Section \ref{sec:input} ($f_{\rm PDCZ}=0.0075$ and
   $f_{\rm CE}=0$) our sequences are able to reproduce several key
   observables of AGB and post-AGB stars in our Galaxy and in the
   Magellanic Clouds. We also compare the IFMRs, HFC growth, and carbon enrichment of our sequences during the TP-AGB with
   those of  state-of-the-art TP-AGB models. This comparison
   shows that our models are good representatives of modern AGB
   stellar evolution models.

   \begin{figure}
            \includegraphics[width=\hsize]{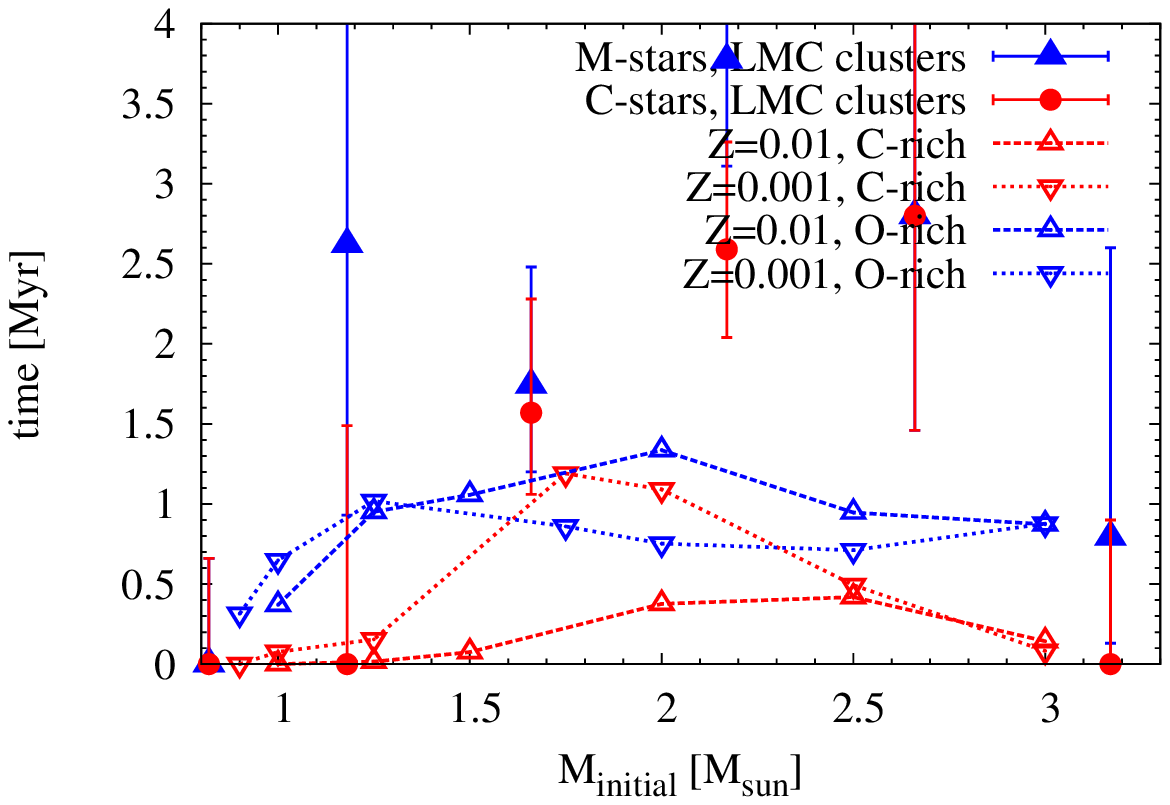}

            \includegraphics[width=\hsize]{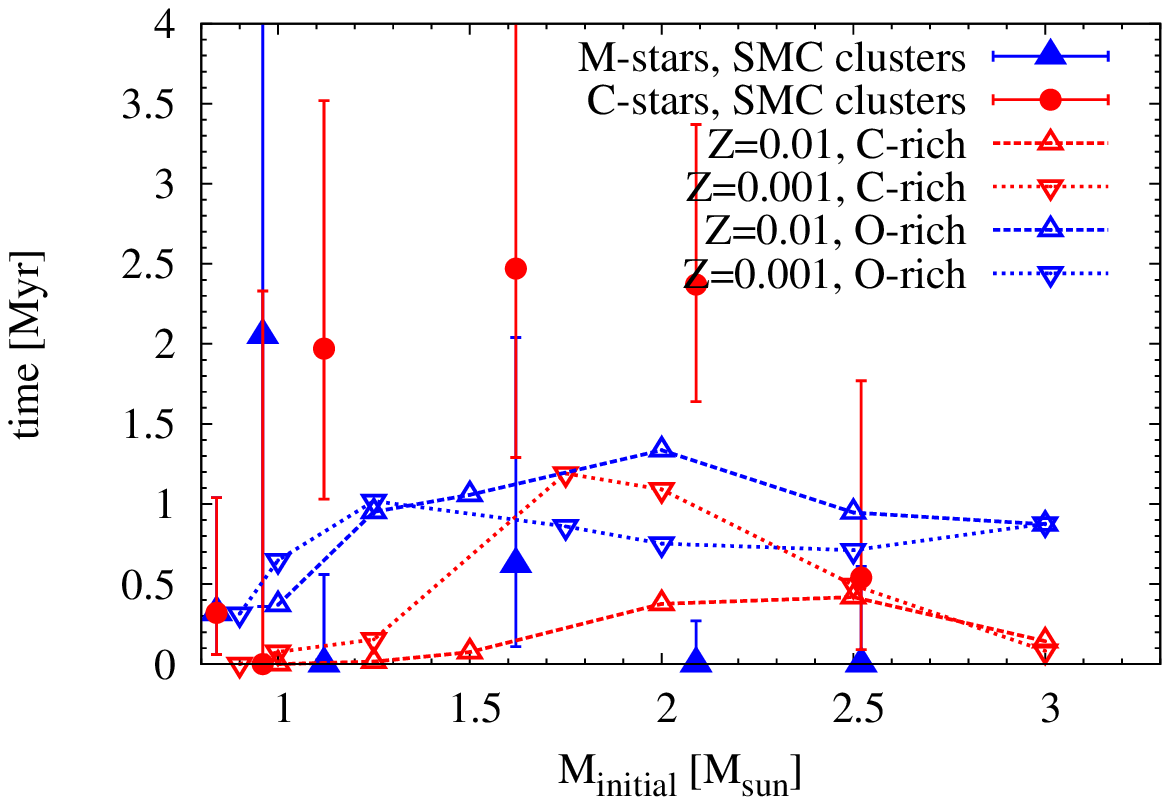}
            \caption{Lifetimes of our $Z_0=0.01$ and $Z_0=0.001$ sequences as C and M
              stars on the AGB as compared with the timescales derived by
              \cite{2007A&A...462..237G} from the Magellanic Clouds. In the case of M-type stars, we computed the
              lifetime of the star as a O-rich object with $M_{\rm Bol}<-3.6$
              to be consistent with the values presented
              by \cite{2007A&A...462..237G}. This is  why the values
              plotted here do not agree with the $\tau_{TP-AGB(M)}$ in table
              \ref{tab:ZAMS-TPAGB}.}
         \label{Fig:LMC-SMC}
   \end{figure}

 \begin{figure}
   \centering
   \includegraphics[width=\hsize]{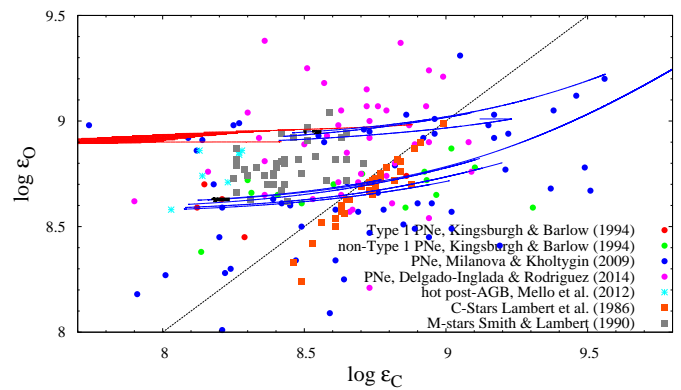}
   \caption{Carbon and oxygen abundances of AGB and post-AGB objects
     in the Galactic disk as compared with the predictions of the
     $Z_0=0.02$ (upper set) and $Z_0=0.01$ sequences (lower set). Tracks
     in blue lines correspond to those that experience significant
     3DUP, while thick black lines indicate sequences with no
     significant 3DUP. The red line corresponds to the evolution of
     the $4M_\odot$ ($Z_0=0.02$) sequence that experiences both 3DUP and
     hot-bottom burning. Abundances are presented in the customary
     astronomical scale for logarithmic abundances, $\log
     \epsilon_{\rm X} = \log(N_{\rm X}/N_{\rm H})+12$, where $N_{\rm
       X}$ and $N_{\rm H}$ are the number densities of elements $X$
     and $H$, respectively.}
         \label{Fig:Comp-AGB-obs}
   \end{figure}

   \begin{figure}
   \centering
   \includegraphics[width=\hsize]{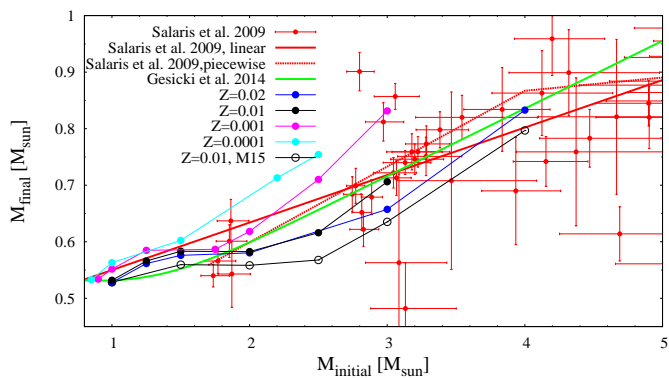}
   \caption{Initial final mass relation of the sequences computed in this
     work and those of \cite{2015ASPC..493...83M} (M15). The semiempirical
     IFMR derived from clusters for solar-like
     metallicities by \cite{2009ApJ...692.1013S}, and also that of
     \cite{2014A&A...566A..48G}, are shown for comparison.}
         \label{Fig:MiMf}
   \end{figure}
   \begin{figure}
   \centering
   \includegraphics[width=\hsize]{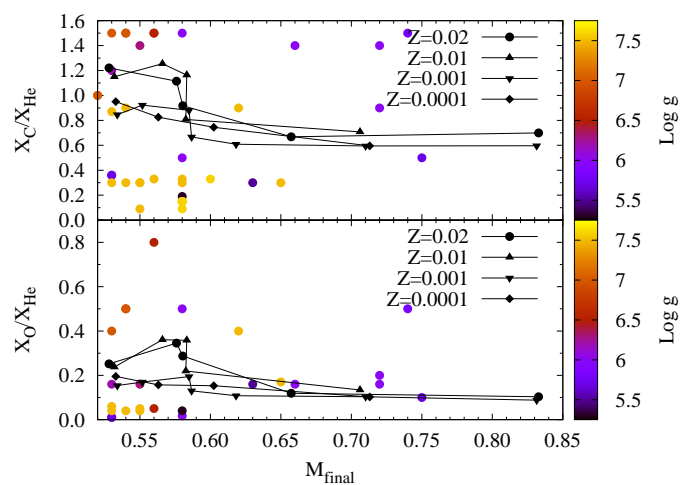}
   \caption{He-, C-, and O- intershell abundances in the post-AGB phase as
     compared with the derived abundances of PG1159 type stars of similar mass
     \citep{Werner2016}. Most of the high $\log g$ stars (see color
     bar) are probably undergoing gravitational settling, and thus the high He
     abundances might not reflect the original intershell composition.}
         \label{Fig:PG1159}
   \end{figure}

   On the AGB most of our sequences undergo efficient 3DUP events and
    carbon enrichment of the envelope. Only the lower mass models avoid a
    carbon enrichment of the envelope. As a consequence of efficient 3DUP events,
   during the TP-AGB most of our models become C-rich; i.e., $N_{\rm C}/N_{\rm O}>1$;
   see table \ref{tab:ZAMS-TPAGB}.  The lower limit for the formation of
   carbon stars is dependent on the initial metallicity of the sequences, being at
   $M_i\sim 1.5\, M_\odot$, $ 1.25\, M_\odot$, $ 1\, M_\odot$ and $0.85\,
   M_\odot$ for $Z_0=0.02$, $0.01$, $0.001$ and $0.0001$, respectively.
   Fig. \ref{Fig:LMC-SMC} shows that our sequences reproduce the range of
   masses for the formation of carbon stars inferred from the study of globular
   clusters in the Magellanic Clouds \citep{2007A&A...462..237G}. Almost no
   carbon stars are predicted at low masses ($\lesssim 1.25 M_\odot$, the exact
   value depending on initial metallicity) because of the lack of 3DUP events and
   almost  no carbon stars are predicted at higher masses ($\gtrsim 3 M_\odot$).

   When AGB models become C rich, they become colder owing to the presence of
   C-rich molecules, as originally shown by \cite{2002A&A...387..507M}; see
   also
   \cite{2009A&A...508.1343W,2010MNRAS.408.2476V,2012ApJ...747....2L,2014ApJ...784...56C}
   for discussions on the impact of C-rich molecular opacities in full stellar
   evolution models.  Because of larger radii and different mass-loss rates of
   carbon stars, mass loss is strongly increased. This causes our models to undergo
   only a few more thermal pulses as C-rich stars before most of the H-rich
   envelope is removed and they evolve away from the AGB. This is in agreement
   with observational data of the C/O ratio of PNe; see
   Fig. \ref{Fig:Comp-AGB-obs}. If many thermal pulses were to follow the carbon-star
   formation, planetary nebulae C/O ratio would cover a much larger range; see \cite{2005ARA&A..43..435H} for a similar discussion. This shows that
   our models depart from the TP-AGB at the right time in terms of carbon
   enrichment, which gives us confidence in the accuracy of the
   post-AGB models.  In addition, our models also reproduce the C/O ratios
   observed  in M- and C-type stars
   \citep{1986ApJS...62..373L,1990ApJS...72..387S} and post-AGB objects
   \citep{1994MNRAS.271..257K,2009AstL...35..518M,2012A&A...543A..11M,2014ApJ...784..173D}
   of the Galactic disk; see Fig. \ref{Fig:Comp-AGB-obs}.

   Third dredge up is a key process in the shaping of the IFMR
   \citep{2009ApJ...692.1013S} and the core-luminosity relation
   \citep{1998A&A...340L..43H, 1999A&A...344..617M}.  Both the
   core-luminosity relation and the IFMR set the timescales of the
   post-AGB remnants (see Appendix \ref{app:physics}). A proper
   modeling of 3DUP episodes becomes of utmost importance for the
   accuracy of post-AGB sequences. Fig. \ref{Fig:MiMf} shows that our
   sequences reproduce the semiempirical IFMR
   (e.g., \citealt{2009ApJ...692.1013S,2014A&A...566A..48G}). This
   implies that our post-AGB models of a given mass $M_f$  descend
   from  reliable progenitor models. Models with high
    CBM efficiencies at the the bottom of the convective envelope
   (e.g., \citealt{2009A&A...508.1343W} and M15) produce final masses
   that are too low at $M_i\sim 3M_\odot$.

   Our models also reproduce  to a good extent the range of He-,
   C-, and O- intershell abundances of AGB stars, as determined from
   the observations of PG1159 stars (Fig.  \ref{Fig:PG1159}; see
   \citealt{2006PASP..118..183W, Werner2016}).  This is an important
   result, as PG1159 stars are the only way to constrain the value of
   $f_{\rm PDCZ}$, which does affect the intensity of the He flashes
   and the consequent 3DUP episodes.  Fig.  \ref{Fig:PG1159} shows
   that the choice of $f_{\rm PDCZ}=0.0075$ allow our models to
   reproduce the O-rich abundances of PG1159 stars; see
   \cite{1999A&A...349L...5H} and \cite{2005ARA&A..43..435H} for a
   detailed discussion.

  \begin{figure}
   \centering
   \includegraphics[width=\hsize]{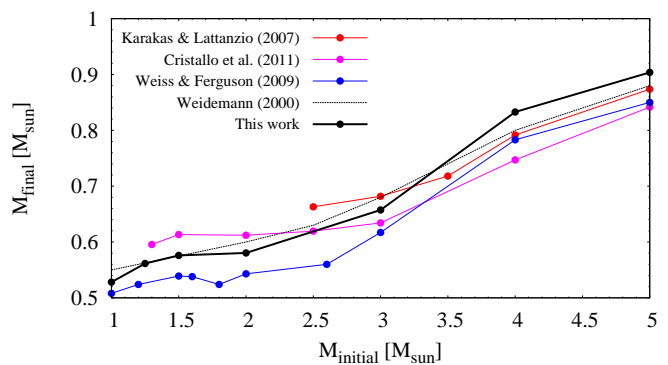}

   \includegraphics[width=\hsize]{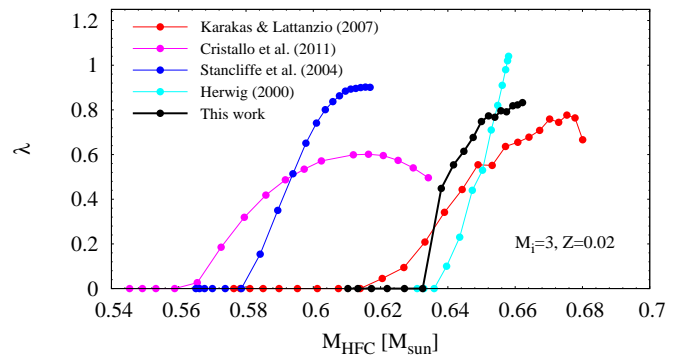}
   \caption{{\it Upper panel:} IFMR of the models presented here for
     $Z_0=0.02$ compared with the results from  state-of-the-art AGB model grids
     available in the literature
     \citep{2007PASA...24..103K,2011ApJS..197...17C,2009A&A...508.1343W} and
     the IFMR of \cite{2000A&A...363..647W}. {\it Lower panel:} Third dredge
     up efficiencies ($\lambda$) during the TP-AGB for different models of a
     $3M_\odot$ and $Z_0=0.02$
     \citep{2000A&A...360..952H,2004MNRAS.352..984S,2007PASA...24..103K,2011ApJS..197...17C}.}
         \label{Fig:Comp-Others}
   \end{figure}
  \begin{figure}
   \centering
   \includegraphics[width=\hsize]{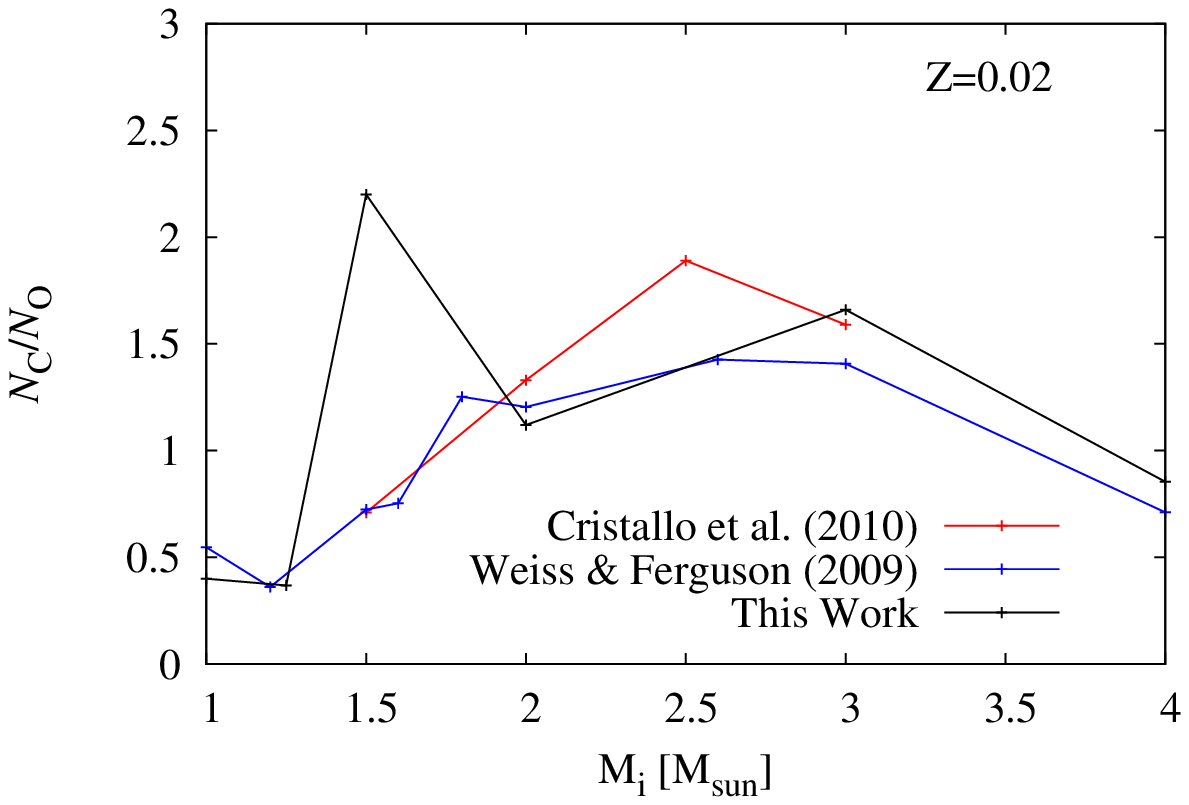}

   \includegraphics[width=\hsize]{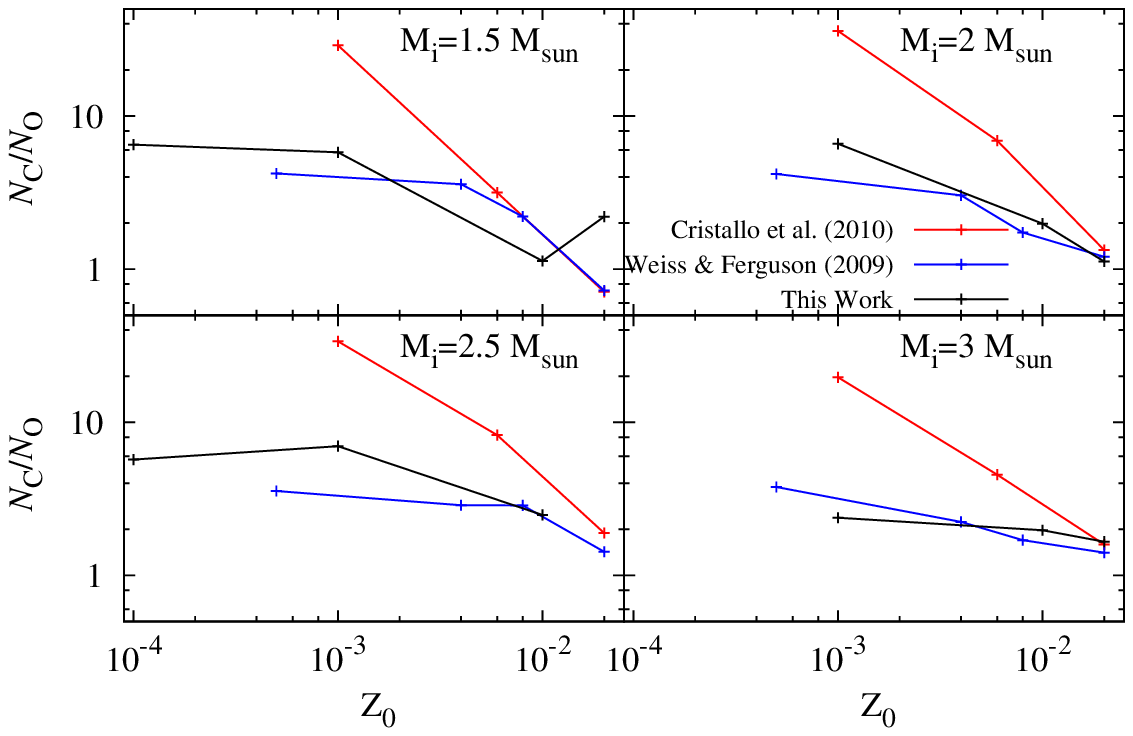}
   \caption{C/O ratios at the end of the TP-AGB of the present models as
     compared with results available in the literature \citep{2009A&A...508.1343W,2011ApJS..197...17C}.}
         \label{Fig:Comp-Others-CO}
   \end{figure}
   The reproduction of the key AGB and post-AGB observables related to
   the HFC growth and  carbon pollution during the TP-AGB
   (Figs. \ref{Fig:LMC-SMC}, \ref{Fig:Comp-AGB-obs}, \ref{Fig:MiMf}
   and \ref{Fig:PG1159}) give us confidence that our post-AGB 
     sequences begin with accurate post-AGB structures.
   Yet, it is possible to wonder to which extent our AGB models, and
   consequently the following post-AGB phase, are representative of
   our current understanding of stellar evolution. The upper panel of
   Fig. \ref{Fig:Comp-Others} shows the IFMR of our models compared
   with the IFMR of available  state-of-the-art AGB models
   \citep{2007PASA...24..103K,2009A&A...508.1343W,2011ApJS..197...17C}. The
   final masses of our models are well within those predicted by
   available grids. The lower panel of Fig. \ref{Fig:Comp-Others}
   shows the evolution of the 3DUP efficiency\footnote{$\lambda=\Delta
     M_{\rm 3dup}/\Delta M_{\rm inter}$; where $\Delta M_{\rm inter}$
     is the increase of the mass of the HFC during the previous
     interpulse phase and $\Delta M_{\rm 3dup}$ is the decrease of the
     mass of HFC during the 3DUP event.} $\lambda$ as a function of
   the mass of the HFC for the benchmark case of a $M_i=3 M_\odot$,
   $Z_0=0.02$ sequence.  Also in this case our model shows a behavior
   well within the spread of the predictions of different
    state-of-the-art AGB models. This spread is mostly due to the
   different treatments of boundary mixing processes in different
   stellar evolution codes and is representative of the present
   uncertainties. In particular, our model is in good agreement with
   the predictions by \cite{2007PASA...24..103K}.

   Fig. \ref{Fig:Comp-Others-CO} shows the predictions for the C/O
   ratio of our post-AGB models compared with the C/O-ratios at the
   end of the AGB for available grids. As shown in the upper panel of
   Fig. \ref{Fig:Comp-Others-CO}, our models predict very similar C/O
   ratios to the models of \cite{2009A&A...508.1343W} and
   \cite{2011ApJS..197...17C} at solar metallicities. The only
   exception is the high C/O ratio of our $M_i=1.5 M_\odot$ ($Z_0=0.02$)
   model. The high C/O ratio of this model is explained by the
   occurrence of a final thermal pulse at the very end of the
   TP-AGB. In those cases, the C dredged up to the surface is diluted
   into a significantly smaller mass of H, leading to a  higher
     surface carbon abundance.  While the three sets of models
   predict rather similar values at solar metallicities, they differ at
   lower metallicities. In particular, the models of
   \cite{2011ApJS..197...17C} predict very high C/O ratios of $N_{\rm
     C}/N_{\rm O}>20$ for $Z_0=0.001,$ while our models and also those
   of \cite{2009A&A...508.1343W} predict a more moderate  carbon
     enrichment. The values of $N_{\rm C}/N_{\rm O}>10$ are in
   significant disagreement with the observed C/O ratios of PNe
   (Fig. \ref{Fig:Comp-AGB-obs}) and also with the recent study of
   \cite{2015MNRAS.452.3679V}. As in the case of the comparisons of
   Fig. \ref{Fig:Comp-Others}, we see that the calibration and choice
   of physics described in Section \ref{sec:input} leads to TP-AGB
   properties in good agreement with independent works.

   We have shown in this section that the models we present here  are able to reproduce several observed AGB and post-AGB
   properties both of the Galactic disk and the Magellanic Clouds; see Figs. \ref{Fig:PG1159}, \ref{Fig:MiMf},
   \ref{Fig:Comp-AGB-obs}, and \ref{Fig:LMC-SMC}. In addition, when
   compared with other state-of-the-art AGB models they usually
   predict properties that are in between those predicted by the
   available  state-of-the-art AGB models; Figs. \ref{Fig:Comp-Others} and \ref{Fig:Comp-Others-CO}. Our
   models are good representatives of the predictions of modern AGB
   computations.

\section{Post-AGB evolution}  %
   \label{sec:post-AGB}

  \begin{table*} 
\caption{ Main post-AGB properties of the H-burning
  sequences.}
\label{tab:pAGB}
    \centering 
\begin{tabular}{cccccccccc} 
\hline\hline 
$M_i$ & $M_f$&
      $\tau_{tr}$ & $ \tau_{cross}$ & $X_{\rm H}$ & $X_{\rm He}$ &
      $X_{\rm C}$ & $X_{\rm N}$ & $X_{\rm O}$ & $\Delta M_{\rm
        env}^{\rm winds}/\Delta M_{\rm env}^{\rm total}$
      \\ $[M_\odot]$ & $[M_\odot]$ & [kyr] & [kyr] & & & & & &
      \\ \hline \multicolumn{10}{c}{$Z_0=0.02$}\\  \hline 1.00 & 0.5281 & 9.14 & 24.9 &
      .671 &.309 &.287E-02 &.181E-02 &.952E-02 &0.167 \\ 1.25 & 0.5615
      & 4.09 & 5.97 & .673 &.307 &.265E-02 &.203E-02 &.956E-02 &0.208
      \\ 1.50 & 0.5760 & 3.39 & 4.49 & .637 &.308 &.281E-01 &.218E-02
      &.170E-01 &0.223 \\ 2.00 & 0.5804 & 2.27 & 1.99 & .661 &.309
      &.985E-02 &.271E-02 &.117E-01 &0.244 \\ 3.00 & 0.6573 & 1.21 &
      .378 & .645 &.321 &.133E-01 &.322E-02 &.107E-01 &0.327 \\ 4.00 &
      0.8328 & .587 & .499E-01 & .627 &.333 &.622E-02 &.159E-01
      &.102E-01 &0.569 \\ \hline \multicolumn{10}{c}{$Z_0=0.01$}\\ \hline 1.00 & 0.5319 &
      36.1 & 63.0 & .701 &.289 &.133E-02 &.105E-02 &.474E-02 &0.201
      \\ 1.25 & 0.5660 & 4.61 & 9.30 & .586 &.310 &.733E-01 &.990E-03
      &.248E-01 &0.154 \\ 1.50 & 0.5832 & 3.10 & 4.22 & .700 &.285
      &.503E-02 &.117E-02 &.592E-02 &0.173 \\ 2.00 & 0.5826 & 2.43 &
      2.40 & .685 &.292 &.109E-01 &.144E-02 &.734E-02 &0.188 \\ 2.50 &
      0.6160 & 1.67 & 1.22 & .675 &.300 &.129E-01 &.163E-02 &.691E-02
      &0.216 \\ 3.00 & 0.7061 & 1.08 & .339 & .684 &.297 &.850E-02
      &.171E-02 &.576E-02 &0.293 \\ \hline
      \multicolumn{10}{c}{$Z_0=0.001$}\\ \hline 0.900 & 0.5340 & 9.89 & 67.3 & .731 &.268
      &.111E-03 &.136E-03 &.468E-03 &0.052 \\ 1.00 & 0.5517 & 4.49 &
      10.4 & .684 &.284 &.262E-01 &.115E-03 &.528E-02 &0.060 \\ 1.25 &
      0.5849 & 2.46 & 3.44 & .649 &.296 &.444E-01 &.166E-03 &.101E-01
      &0.074 \\ 1.75 & 0.5867 & 1.82 & 2.14 & .549 &.332 &.872E-01
      &.371E-03 &.181E-01 &0.078 \\ 2.00 & 0.6182 & 1.57 & 1.28 & .661
      &.294 &.345E-01 &.185E-03 &.698E-02 &0.097 \\ 2.50 & 0.7101 &
      .937 & .318 & .642 &.297 &.463E-01 &.532E-03 &.882E-02 &0.195
      \\ 3.00 & 0.8314 & .763 & .117 & .691 &.278 &.767E-02 &.168E-01
      &.428E-02 &0.275 \\ \hline \multicolumn{10}{c}{$Z_0=0.0001$}\\  \hline 0.850 & 0.5328 & 4.63
      & 42.5 & .459 &.346 &.161 &.142E-04 &.333E-01 &0.038 \\ 1.00 &
      0.5631 & 3.02 & 5.44 & .677 &.286 &.310E-01 &.147E-04 &.590E-02
      &0.038 \\ 1.50 & 0.6024 & 1.76 & 1.98 & .619 &.308 &.557E-01
      &.191E-03 &.114E-01 &0.049 \\ 2.20 & 0.7130 & .951 & .377 & .641
      &.297 &.310E-01 &.172E-01 &.847E-02 &0.116 \\ 2.50 & 0.7543 &
      .715 & .184 & .566 &.326 &.603E-01 &.205E-01 &.141E-01 &0.189
      \\ \hline
\multicolumn{10}{p\textwidth}{$M_i$: Initial mass of the model (at ZAMS). $M_f$: Final
  mass of the star. $\tau_{\rm tr}$: Timescale from the end of the AGB
  (taken at $M_{\rm env}=0.01 M_\star$) to the moment in which $\log
  T_{\rm eff}= 3.85$. $ \tau_{cross}:$ Timescale from the moment in
  which $\log T_{\rm eff}= 3.85$ to the point of maximum effective
  temperature. $X_{\rm H}$, $X_{\rm He}$, $X_{\rm C}$, $X_{\rm N}$,
  and $X_{\rm O}$: Surface abundances H, He, C, N, and O of the
  post-AGB models. $\Delta M_{\rm env}^{\rm winds}$ and $\Delta M_{\rm
    env}^{\rm total}$: Reduction of the H-rich envelope ($M_{\rm
    env}$) from $\log T_{\rm eff}= 3.85$ to the point of maximum
  $T_{\rm eff}$ owing to winds and  the combined effect of winds
  and H-burning, respectively.}
\end{tabular} \end{table*}

\subsection{Description of the    results} 
\begin{figure}
    \centering \includegraphics[width=\hsize]{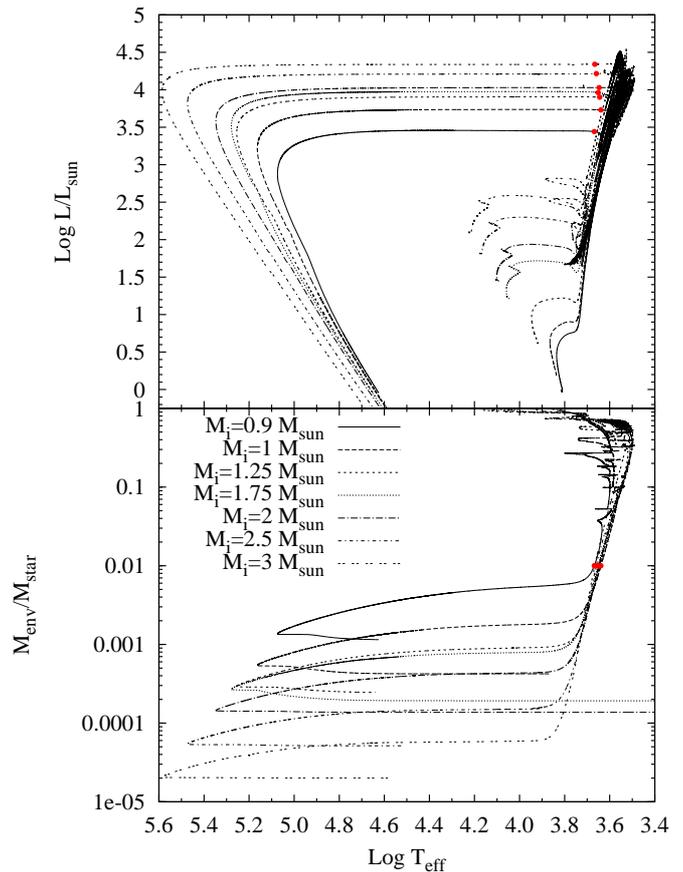} \caption{Upper
      panel: HR diagram of our $Z_0=0.001$ sequences that departed from the AGB as
      H-burners. Lower panel: Mass of the H-rich envelope as a function of
      $\log T_{\rm eff}$ for the same sequences. Red dots indicate the moment in
      which $M_{\rm env}/M_{\rm star}=0.01$, which has been used to define the
      beginning of the
      post-AGB.}  \label{Fig:EnvMass} \end{figure} 

\begin{figure*} \includegraphics[width=\textwidth,clip,clip]{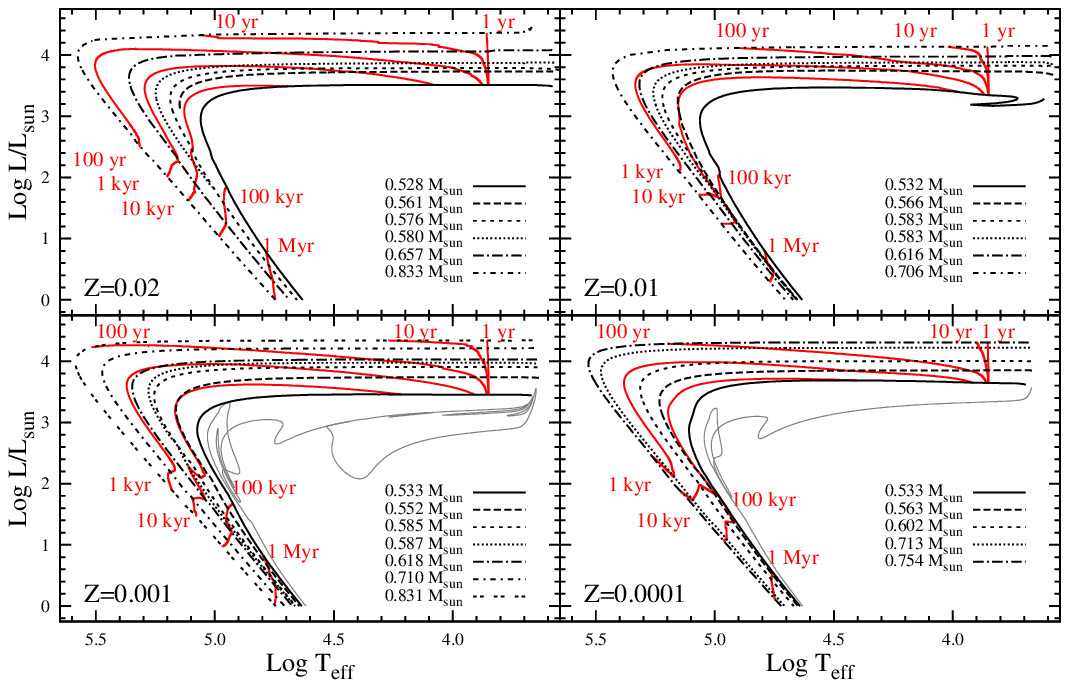} \caption{HR diagrams
      of the H-burning post-AGB sequences we computed for different
      masses and initial metallicities. Tracks are presented from the
      beginning of the post-AGB phase when the H-rich envelope drops below
      $M_{\rm env}=0.01 M_\star$ to the moment in which the star has already
      entered its white dwarf cooling sequence at $L_\star=L_\odot$. At that
      point gravitational settling should have already started to turn
      post-AGB stars into DA-WDs, a process not included in the present
      computations. Red lines indicate computed isochrones for different ages
      since the zero point defined at $\log T_{\rm eff}=3.85$. Thin gray lines
      in the $Z_0=0.001$ and $Z_0=0.0001$ panels correspond to the evolution of
      the $M_i=0.8M_\odot$ He-burning sequences shown in
      Fig. \ref{Fig:He-burners} ($M_f=0.4971 M_\odot$ and $0.5183 M_\odot$,
      respectively). } \label{Fig:pAGB} \end{figure*}

  In our models, the departure from the AGB occurs as a consequence of
  steady winds reducing the mass of the H-rich envelope $M_{\rm env}$
  \citep{1979A&A....79..108S}.  Consequently, this transition occurs gradually
  and the definition of the beginning of the post-AGB phase (i.e., end of the
  AGB) is somewhat arbitrary. The mass of the envelope as function of
  effective temperature is shown in Fig. \ref{Fig:EnvMass} for our H-burning
  sequences with $Z_0=0.001$.  As soon as the mass of the envelope becomes
  comparable to the mass of the core ($M_{\rm env}/M_{\rm star}=0.5\mbox{--}0.1,$
  depending on mass; see Fig. \ref{Fig:EnvMass}), models move to the blue with
  decreasing envelope mass. First, big changes in the envelope mass lead only
  to a modest increase in temperature. This phase lasts until the mass of the
  envelope becomes a few times (between three and five times in the sequences shown
  in Fig. \ref{Fig:EnvMass}) larger than the final mass of the white dwarf
  envelope. After this point, small changes in the envelope mass lead to big
  changes in the effective temperature of the model. Most models depart from
  the AGB region in the HR diagram during the first, slower  stage. Consequently, there is no natural definition of the end of the
  AGB. We have defined the beginning of the post-AGB phase as the
  moment in which $M_{\rm env}/M_{\rm star}=0.01$. At this moment, models have
  already moved significantly to the blue, which is true at all masses and
  metallicities. This choice defines the end of the AGB in a homogeneous way
  independently from their mass and metallicities, and is based on the
  underlying physical reason behind the departure from the AGB.

  Fig. \ref{Fig:pAGB} displays the evolution of our H-burning post-AGB
  sequences in the theoretical HR diagram for different metallicities and
  masses.  In works dealing with the evolution of post-AGB stars and CSPNe
  \citep{1989IAUS..131..391R, 1990fmpn.coll..355S,1994ApJS...92..125V,
    2004A&A...423..995M,2009A&A...508.1343W}, it is customary to define two
  different timescales, $\tau_{tr}$ and $ \tau_{cross}$, for the discussion of
  the post-AGB evolution. The quantity  $\tau_{tr}$ corresponds to the duration of the
  early post-AGB evolution when AGB-like winds might still be important and
  $T_{\rm eff}$ does not depend strongly on $M_{\rm env}$.  The value $ \tau_{cross}$, in
  turn, gives the timescale of the later post-AGB evolution when there is a
  tight $T_{\rm eff}-M_{\rm env}$ relation, from the end of the early phase to
  the point of maximum effective temperature.  Our simulations show that
  around $\log T_{\rm eff}\sim 3.85$ all our sequences have started the fast
  part of the post-AGB evolution. In addition, at these temperatures (and
  beyond) winds play only a secondary role in setting the timescales. This
  makes timescales in the second phase more reliable than in the early
  post-AGB phase. Splitting the post-AGB timescale in $\tau_{cross}$ and
  $\tau_{tr}$ allows $\tau_{cross}$ to be a useful and reliable
  quantity. The quantity $\tau_{cross}$ is then unaffected by our lack of understanding of
  the early post-AGB winds and the absence of a clear end of the TP-AGB. We
  define $\tau_{tr}$ as the time from the end of the AGB (taken at $M_{\rm
    env}=0.01 M_\star$) to the moment in which $\log T_{\rm eff}= 3.85$, while
  $ \tau_{cross}$ is the timescale from $\log T_{\rm eff}= 3.85$ to the point
  of maximum effective temperature. The value of $ \tau_{cross}$ is almost
  independent of the precise definition and would have been practically the
  same if the initial point were set at $\log T_{\rm eff}= 4$ as in
  \cite{1994ApJS...92..125V} and \cite{2009A&A...508.1343W} or if we had adopted the
  definition of \cite{1995A&A...299..755B} \footnote{
    \cite{1995A&A...299..755B} defines the end of the transition stage at the
    point where the pulsation period $P=50$d, which also sets the zero age for
    their post-AGB tracks. This definition corresponds to a point in the
    HR diagram where the post-AGB object has $\log T_{\rm eff}= 3.78
    \mbox{--} 3.90$, depending on mass.}.

\begin{figure*}
   \centering
   \includegraphics[width=\textwidth]{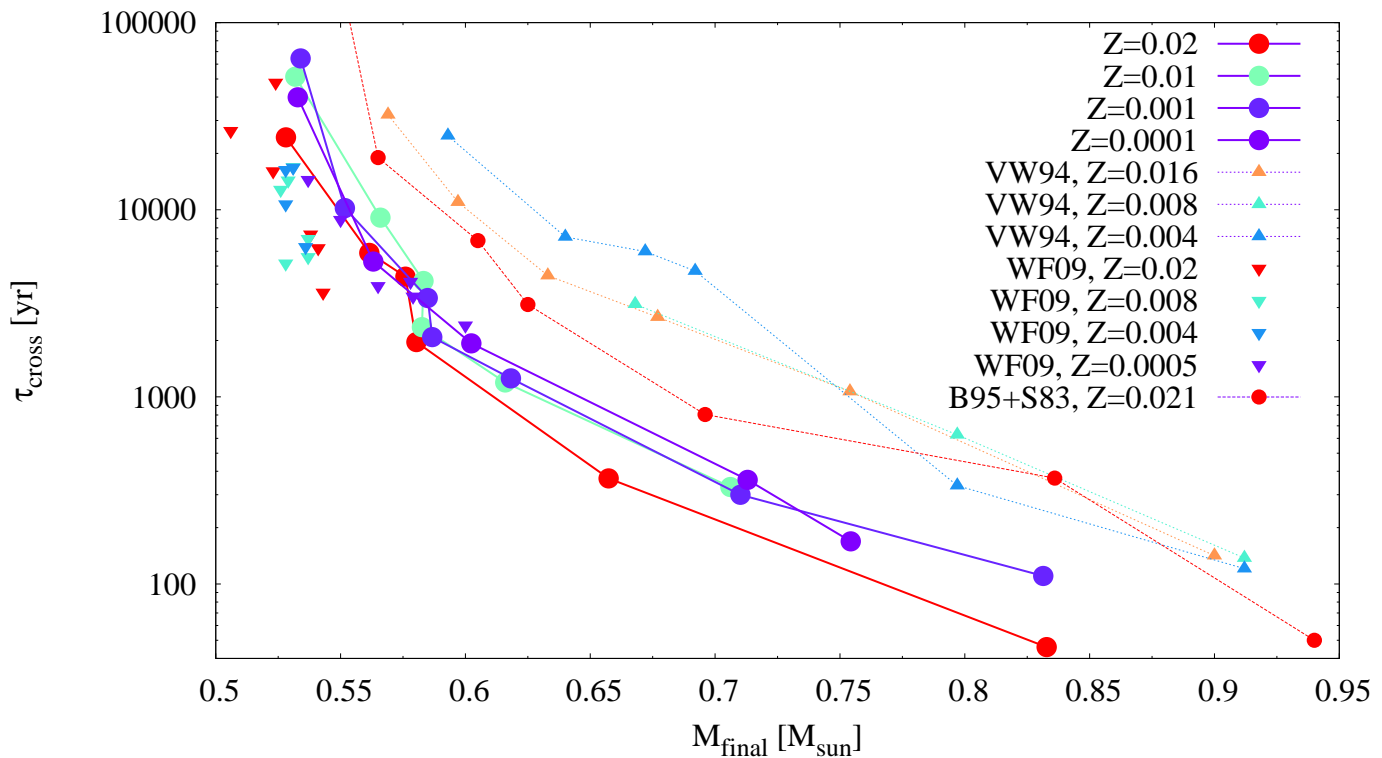}
   \caption{Crossing timescales from the zero point, set at $\log
     T_{\rm eff}=4$ to allow comparisons with all previous works, to
     the point of maximum effective temperature in the HR diagram; i.e., the so-called knee of the evolutionary tracks shown in
     Fig. \ref{Fig:pAGB}. Timescales are shown for all the sequences
     presented  and the H-burning post-AGB sequences
     available in the literature, i.e.,  \cite{1994ApJS...92..125V}
     (VW94); \cite{1983ApJ...272..708S,1995A&A...299..755B} (B95+S83);     and \cite{2009A&A...508.1343W} (WF09). Crossing timescales of the
     new sequences (this work and also \citealt {2009A&A...508.1343W})
     are much shorter than the crossing timescales of the H-burning
     models of
     \cite{1983ApJ...272..708S,1994ApJS...92..125V,1995A&A...299..755B}. The
     sudden decrease in $\tau_{\rm cross}$ at $M_f\sim 0.58M_\odot$
     ($Z_0=0.02$, 0.01 and 0.001) can be traced back to differences in the
     previous evolutionary history of the models  (see text). Because of slightly different definitions of the zero point, the
     numbers do not completely agree with the value of $\tau_{\rm
       cross}$ given in Table \ref{tab:pAGB}.}
         \label{Fig:CrossTimes}
   \end{figure*}

  Table \ref{tab:pAGB} and Fig. \ref{Fig:pAGB} shows the main post-AGB
  properties of the H-burning sequences computed in this work.  We emphasize the
  extreme mass dependence of the post-AGB timescales. While higher mass models
  ($\gtrsim 0.7\ M_\odot$) require only a few hundreds of years to cross the
  HR diagram and only thousands of years to fade two orders of magnitude,
  lower mass models ($\lesssim 0.55\ M_\odot$) require more than 10 kyr to
  cross the HR diagram and more than 100 kyr to decrease their luminosity only
  by an order of magnitude.  The predicted post-AGB timescales do not depend
  strongly on the initial metallicity of the population (i.e., iron content).
  A sudden decrease in timescales, around $M_{\rm f}\sim 0.58 M_\odot$ for
  most metallicities, is apparent in both Table \ref{tab:pAGB} and Fig
  \ref{Fig:CrossTimes}.  This corresponds to the transition from models that
  did, and did not, undergo a HeCF in the previous evolution, i.e., models
  with $M_{\rm i}= 1.5$ and $2 M_\odot$ for $Z_0=0.02,\ 0.01$, and to $M_{\rm
    i}= 1.25$ and $1.75 M_\odot$ for $Z_0=0.001$.  These sequences end with
  similar $M_{\rm f}$ but they reached the AGB with different HFC masses
  ($M_c^{1TP}$; Table \ref{tab:ZAMS-TPAGB}). Consequently,  the time spent on the
  AGB and the chemical and thermal structure of the core at the end of the AGB
  are not the same.  As discussed in the Appendix \ref{app:physics}, the
  thermal structure of the core and  CNO enrichment of the envelope play a
  key role in setting the properties of the post-AGB models.  As a result of a
  different 3DUP history, the composition of the envelope is
  different, with models on the high-mass side of the transition showing
  more efficient 3DUP. Models on the high-mass side of the
  transition  have higher luminosities and less massive  H envelopes. As the
  timescale for the crossing of the HR diagram ($ \tau_{cross}$ ) is given by
  the speed at which the envelope is depleted, both higher H-burning rates and
  smaller initial H envelopes lead to shorter timescales for models with
  $M_f\gtrsim 0.58 M_\odot$.

   We can estimate the so-called transition times ($\tau_{\rm tr}$, Table
   \ref{tab:pAGB}) immediately after the departure form the AGB. Table
   \ref{tab:pAGB} shows that $\tau_{tr}$ has a very steep decrease for
   remnant masses between $\sim 0.53M_\odot$ and $\sim 0.58M_\odot$
   and a flatter decrease at higher masses. The value of $\tau_{\rm
     tr}$ is both sensitive to the precise definition of the end of
   the AGB and to the intensity of winds during the early post-AGB
   phase, which are  poorly known. Changing the definition of the
   end of the AGB from $0.01 M_\star$ to $0.007 M_\star$ can change
   $\tau_{\rm tr}$ by more than $50\%$. With this in mind the value of
   $\tau_{\rm tr}$ predicted by the new models goes from $\tau_{\rm
     tr} \sim 2\mbox{--}7$ kyr at $M_f\sim 0.53 M_\odot$ to $\tau_{\rm
     tr}\sim 0.5\mbox{--}1$ kyr $M_f\gtrsim 0.7 M_\odot$ with exact values
   depending on the precise definition of the end of the AGB and
    the metallicity.

   In light of the fact that no effort was made to control the phase at which
   the models depart from the AGB, we stress that a big
   majority of our sequences depart from the AGB, and evolve through the
   post-AGB, as H-burning models. Only the very low-mass and low metallicity
   models ($0.8 M_\odot$ and $Z_0=$0.001, 0.0001) evolved away from the AGB as
   He-burning models. In addition, one sequence underwent a late thermal pulse
   (LTP; \citealt{2001Ap&SS.275....1B}) in the post-AGB evolution
   ($1.5M_\odot$, $Z_0=$0.001) and two other sequences underwent a very late
   thermal pulse (VLTP, \citealt{2001Ap&SS.275....1B}) already on the white
   dwarf cooling track ($1.25$ and $2 M_\odot$ models with $Z_0=0.02$).

\begin{figure}
   \centering
   \includegraphics[width=\hsize]{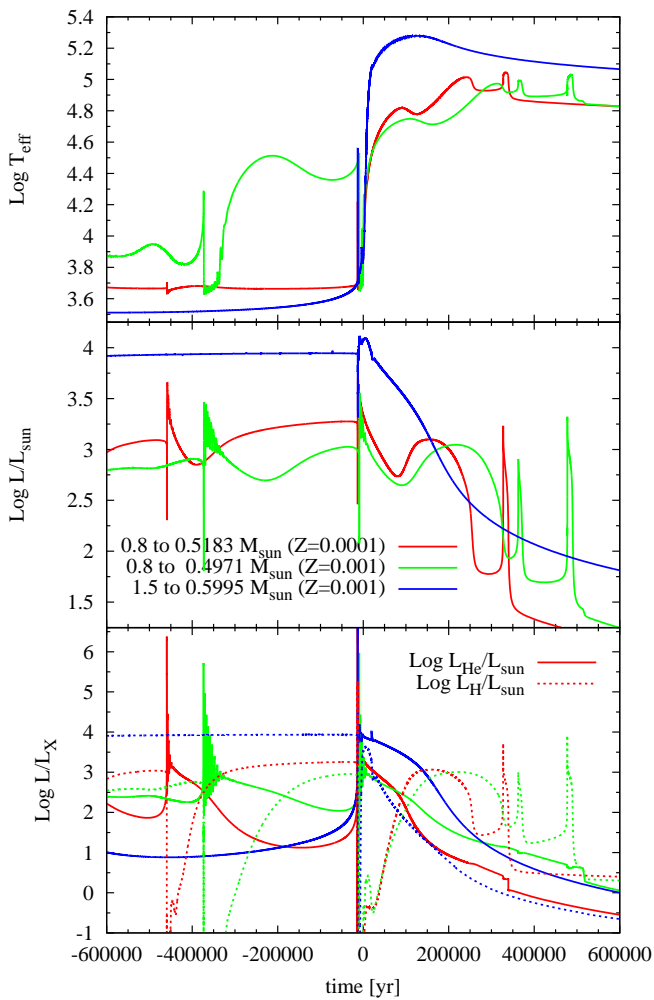}
   \caption{Evolution of $\log T_{\rm eff}$ (upper panel) and $\log L/L_\odot$
     (lower panel) during the post-AGB phase and the end of the TP-AGB. The
     zero point of the x-axis was taken at the last time the sequence had
     $\log T_{\rm eff}=3.85$ before evolving toward the white dwarf cooling
     phase. Time is shown in years for the $M_i=0.8M_\odot$ sequences, but it has
     been multiplied by a factor 10 for the  $M_i=1.5M_\odot$ sequence to allow
     proper visualization.}
         \label{Fig:He-burners}
   \end{figure}

   Fig. \ref{Fig:He-burners} shows the evolution of $\log T_{\rm eff}$ and
   $\log L/L_\odot$ of the only three He-burning sequences computed in this
   work.  The time lapse from\footnote{Taken at the last time the star had
     this temperature before evolving to much higher temperatures.} $\log
   T_{\rm eff}=3.85$ to the point of maximum effective temperature is of $\sim
   487$ kyr ($\sim 334$ kyr) for the $0.4971 M_\odot$ ($0.5183 M_\odot$)
   sequence.  These values are much higher than the crossing times of the
   H-burning sequences of lower mass ($\tau_{cross}\sim 20\mbox{--}60$ kyr for
   $M_f\sim 0.53 M_\odot$; table \ref{tab:pAGB}). While we refer to these
   low-mass sequences as He-burners, they do not spend the whole post-AGB as
   He-burners. Fig.  \ref{Fig:He-burners} shows that only in the first $\sim
   100$ kyr the sequences are He-burners ($L_{\rm He}>L_{\rm H}$) while for
   the rest of the post-AGB H-burning becomes dominant again, as is typical
   from the interpulse phase in the TP-AGB. The reignition of the H-shell in
   the post-AGB phase leads to a temporary decrease in $T_{\rm eff}$ as the
   model readjusts to the new structure. Only objects with very low H
   abundances in the envelope can evolve through the whole post-AGB phase as
   He-burners. This is the case for our LTP $M_i=1.5 M_\odot$, $Z_0=0.001$
   sequence, which undergoes a LTP and finally becomes an object with surface
   abundances of
   [H/He/C/N/O/Ne]=[0.036/0.504/0.353/$5.8\times10^{-4}$/0.075/0.029] because of the dilution of the H envelope during 3DUP. Although H is reignited
   immediately after the LTP, it never overtakes the He-burning energy release
   and the sequence remains a He-burner throughout the post-AGB phase (see
   Fig. \ref{Fig:He-burners}). The crossing timescale for this sequence is
   $\tau_{\rm cross}=12.4$ kyr, which is much larger than the crossing timescales of
   H-burning sequences of similar mass and metallicity (3.4 and 2.1 kyr, see
   Tab. \ref{tab:pAGB}). The evolution of very low-mass post-AGB sequences can
   be very involved with even several flashes taking place in the post-AGB
   phase \citep{1995A&A...299..755B}. This is the case of our $0.4971
   M_\odot$ sequence (green curve in Fig. \ref{Fig:He-burners}), which
   undergoes two post-AGB thermal pulses with he first at $t\sim -375 $ kyr in
   Fig.  \ref{Fig:He-burners}. Very low-mass post-AGB sequences also
   experience some sudden enhancements in the H-burning shell just before the
   point of maximum temperature, as the star model contracts toward the white
   dwarf cooling phase; the spikes in $L_{\rm H}$ are just at the point of
   maximum effective temperature; Fig. \ref{Fig:He-burners}.

\subsection{Comparison with previous post-AGB grids}

\begin{figure}
   \centering
   \includegraphics[width=\hsize]{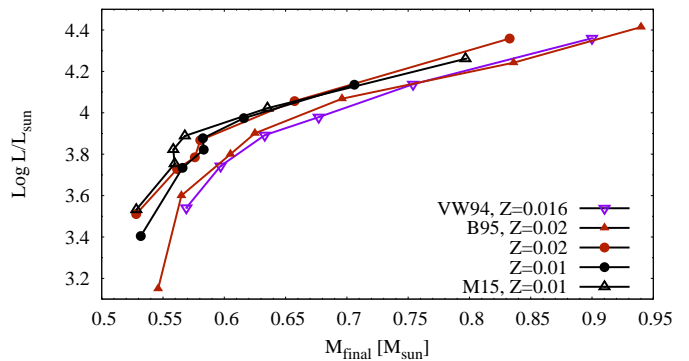}
   \caption{Post-AGB mass-luminosity relation of the old  H-burning models as compared
     with that of the new models we present here and in
     \cite{2015ASPC..493...83M}. There is a much
     higher luminosity for the same remnant mass of the new models. }
         \label{Fig:Core_Lum}
   \end{figure}

The main result of our study is the finding that new models predict
post-AGB timescales ($\tau_{cross}$)  that are three to ten times
shorter than the older models of \cite{1994ApJS...92..125V} and
\cite{1995A&A...299..755B} (see Fig. \ref{Fig:CrossTimes}).  This
result is in agreement with the previous results found by
\cite{2009A&A...508.1343W} in the low-mass range
(Fig. \ref{Fig:CrossTimes}). The agreement is reassuring  because
the results of \cite{2009A&A...508.1343W} were computed with a
different stellar evolution code, but also include a  state-of-the-art
modeling of the TP-AGB.  The new models are  also brighter by about
$\sim 0.1... 0.3$ dex for most remnant masses, as shown in
Fig. \ref{Fig:Core_Lum}.

The speed of the post-AGB evolution of H-burning sequences is set by
the speed at which the H-rich envelope is consumed. Models departing
from the TP-AGB with less massive H-rich envelopes, higher
luminosities, or more intense winds must evolve faster than those with
more massive envelopes, lower luminosities, and less intense
winds. With the exception of  the sequence with the highest
  metallicity and mass ($Z_0=0.02$ and $M_i=4 M_\odot$), post-AGB winds always play a secondary role in the depletion of the H-rich
envelope and  even become negligible at lower metallicities ($\Delta
M_{\rm env}^{\rm winds}/\Delta M_{\rm env}^{\rm total}$, Table
\ref{tab:pAGB}). Yet, mass loss still plays a relevant role, on the
order of $\Delta M_{\rm env}^{\rm winds}/\Delta M_{\rm env}^{\rm
  total}\sim 0.15\mbox{--}0.35$, in setting the exact timescale of the 
  higher metallicity sequences ($Z_0=$0.01, 0.02). The secondary role
of mass loss for the post-AGB timescales implies that the reason for
the fast evolution of the new models must be related to the other two
ingredients that define post-AGB timescales, post-AGB luminosities,
and H-envelope masses.

The reason why post-AGB models have different luminosities or H-rich
envelope masses is  involved and is related to the core
mass-luminosity relation on the TP-AGB.  Modern sequences have updated
microphysics and a better modeling of mixing processes on the AGB.
This leads to very different core-luminosity relations in the AGB and
post-AGB phases, as shown in Fig. \ref{Fig:Core_Lum}. These differences
also produce different masses of the H envelope at the moment of the
departure from the TP-AGB.  As discussed in Appendix \ref{app:physics},
new models depart from the AGB with brighter luminosities and smaller
H-envelope masses, producing a faster post-AGB evolution. In addition,
the efficient 3DUP of most of the sequences affects the properties of
post-AGB models through the  carbon enrichment of the H-rich
envelope and its impact on the IFMR (see also Appendix
\ref{app:physics}).  The fact that the new sequences are able to
reproduce several AGB and post-AGB observables
(Figs.\ref{Fig:LMC-SMC}, \ref{Fig:Comp-AGB-obs}, \ref{Fig:MiMf},
\ref{Fig:PG1159}) not reproduced by the older models implies that
the new models should be preferred over older models.  The
difference in envelope masses,  carbon enrichment, and IFMR explains the much
faster post-AGB evolution that we obtained, as compared with the
older grids
\citep{1983ApJ...272..708S,1994ApJS...92..125V,1995A&A...299..755B};
see Appendix \ref{app:physics} for a detailed discussion.

   \begin{figure}
   \centering
   \includegraphics[width=\hsize]{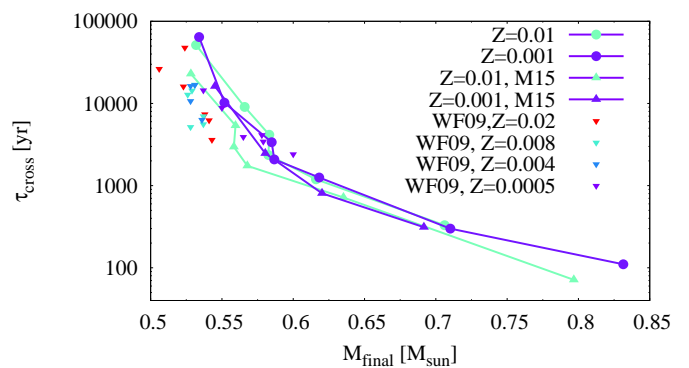}
   \caption{Same as Fig. \ref{Fig:CrossTimes} (although with the zero points
     set at $\log T_{\rm eff}=3.85$), but comparing the  H-burning sequences with those originally computed by \cite{2015ASPC..493...83M}
     under different assumptions of the evolution on the TP-AGB with the
     same code and microphysics. This shows the uncertainty in the computed
     post-AGB crossing timescales due to uncertainties in the treatment of
     3DUP episodes on the TP-AGB.      \cite{2015ASPC..493...83M} models show a better agreement with the
     results of \cite{2009A&A...508.1343W}, which were also computed under the
     assumption of very efficient  CBM processes on the TP-AGB.}
         \label{Fig:Cross_AB}
   \end{figure}

   Fig. \ref{Fig:CrossTimes} shows that the models by
   \cite{2009A&A...508.1343W} predict even shorter timescales than our
   models. This is consistent with the fact that
   \cite{2009A&A...508.1343W} adopted higher  CBM efficiencies,
   $f_{\rm CE}=f_{\rm PDCZ}= 0.016$, during the TP-AGB, and
   consequently have more efficient 3DUP. When our models are computed
   with higher  CBM efficiencies, the predicted post-AGB
   timescales are shorter and very close to the results of
   \cite{2009A&A...508.1343W}. This is shown in
   Fig. \ref{Fig:Cross_AB}, where the value of $\tau_{cross}$ from the
   sequences presented here (computed with $f_{\rm CE}= 0$;
   Table \ref{tab:pAGB}) are compared with the results of
   \cite{2009A&A...508.1343W} and our preliminary computations from
   \cite{2015ASPC..493...83M} (M15 models, computed with $f_{\rm CE}=
   0.13$, see Appendix \ref{app:M15}). Different  CBM
   efficiencies lead to different 3DUP histories. More efficient 3DUP,
   in turn, leads to a higher  carbon enrichment of the envelopes and
   smaller remnant masses $M_f$ for the same initial mass $M_i$. As
   shown in Appendix \ref{app:physics} a higher  carbon enrichment of
   the envelope leads to brighter models (see Fig. \ref{Fig:Core_Lum})
   and smaller post-AGB envelopes, leading to shorter post-AGB
   timescales. In addition, post-AGB models of equal  final mass ($M_f$), 
   coming from larger initial masses ($M_i$), also seem to produce shorter post-AGB timescales (see also Appendix
   \ref{app:physics}). This explains why the post-AGB timescales of
   the models we present here are not as short as those of
   \cite{2009A&A...508.1343W} and M15 models.  \cite{2009A&A...508.1343W} and the M15 models predict final masses
   that are too low for our present understanding of the IFMR,
   especially around $M_i\sim 3M_\odot$
   \citep{2000A&A...363..647W,2009ApJ...692.1013S,2014A&A...566A..48G}. In contrast, the models in table \ref{tab:pAGB} give a much 
   better fit of the IFMR around solar metallicities
   (Fig. \ref{Fig:MiMf}).

   Finally, due to the different definitions of $\tau_{tr}$, it is very
   difficult to directly compare these values with those quoted by
   \cite{1990fmpn.coll..355S}, \cite{1994ApJS...92..125V}, and
   \cite{1995A&A...299..755B}. It can be safely concluded, however, that our
   definition yields $\tau_{tr}$ values that decrease with increasing remnant
   mass, in agreement with both \cite{1990fmpn.coll..355S} and
   \cite{1995A&A...299..755B}, but at variance with the results of
   \cite{1994ApJS...92..125V}, which yield higher values and show no
   significant dependency on the remnant mass.

\section{Discussion}
\label{sec:discussion}

The more accurate description of the AGB and post-AGB observables of the new
models (Figs. \ref{Fig:LMC-SMC}, \ref{Fig:Comp-AGB-obs}, \ref{Fig:MiMf} and
\ref{Fig:PG1159}) makes it reasonable to assume that the new timescales are
more reliable than those of the old post-AGB grids. Post-AGB timescales play a
role in several studies and shorter timescales certainly have an impact
on the conclusions of previous works based on old stellar tracks. In the
following section, we discuss these and speculate on the possible consequences
of the new post-AGB models and their shortcomings.

\cite{2014A&A...566A..48G} determined that the post-AGB evolution predicted by
\cite{1995A&A...299..755B} needed to be accelerated by about a factor $\sim 3$
to reconcile the peaks of the mass distributions of WDs and CSPNe. Even more,
one expects the peak of the mass distribution of CSPNe to lie somewhat below
the peak of the mass distribution of WDs. The timescales of CSPNe are steeply
dependent on mass, making lower mass central stars much more abundant than
more massive central stars. While the higher luminosity of more massive CSPNe makes
them more easily detectable, the difference in luminosities is less
significant than the difference in timescales. One might then take the
factor of $\sim 3$ as a lower limit for the accelerated evolution needed in
\cite{1995A&A...299..755B} post-AGB sequences. As shown in
Fig. \ref{Fig:CrossTimes}, the new post-AGB timescales are $\sim3.5$ and $\sim
8$ times shorter than those computed by \cite{1995A&A...299..755B} in the
range of interest for the peak of the WDs and CSPNe mass distributions
($0.57\lesssim M/M_\odot\lesssim 0.63$). Clearly, the new post-AGB timescales
seem to be in very good agreement with the empirical determinations of
\cite{2014A&A...566A..48G}.

The faster evolutionary timescales and higher luminosities of our
H-burning sequences should have an important impact on the study of
the formation of PNe
\citep{2014AN....335..378S,2014MNRAS.443.3486T}. The low-mass models
of \cite{1983ApJ...272..708S}, which are still in use to complement
the sequences of \cite{1995A&A...299..755B}, show crossing timescales
of $\sim 340$ kyr (0.546 $M_\odot$) and $\sim 20$ kyr (0.565
$M_\odot$). Our H-burning sequences of similar mass and metallicity
($Z_0=0.02$) show crossing timescales about $\sim 3.5$ to $\gtrsim 15$
times faster and even faster in the case of M15 models. The
discrepancy is even larger in the case of the low-mass models of
\cite{1994ApJS...92..125V}. In order to be able to produce PNe,   the central stars need to evolve in less than a few tens of
thousand years. If that is not the case, the circumstellar material
dissipates before the star becomes hot enough to ionize it. This fact,
together with the very long timescales of $\tau_{\rm cross}\gtrsim
100$ kyr of the low-mass models of older grids, leads to the
conventional wisdom that low-mass post-AGB stars ($\lesssim 0.55
M_\odot$) cannot form PNe; see, e.g., \cite{2013ApJ...769...10J,
  2015AJ....149..132B}.  In this context, the new models might help to
explain the existence of single CSPNe with masses lower than $\sim
0.55 M_\odot$ \citep{2010A&ARv..18..471A, Werner2016}. The new models
 should also have an impact on the question of whether single stellar
evolution can form PNe in globular clusters
\citep{2013ApJ...769...10J, 2015AJ....149..132B}. Our H-burning
post-AGB sequences with ages similar to that of globular clusters (9
to 12 Gyr) have values of $\tau_{\rm cross}\sim 25\mbox{--} 70$
kyr. Timescales drop to $\tau_{\rm cross}\sim 5\mbox{--} 10$ kyr for
post-AGB sequences with slightly younger progenitors (5 to 7 Gyr); see
Tables \ref{tab:ZAMS-TPAGB} and \ref{tab:pAGB}. Timescales are even
shorter in the case of the models of \cite{2009A&A...508.1343W} and
M15 (see Appendix \ref{app:M15}). The much shorter timescales of the
new H-burning post-AGB sequences call into question the idea that
single stellar evolution cannot produce PNe in globular clusters.

Some studies of the number of post-AGB stars in old populations like
M32 \citep{2008ApJ...682..319B} and the Galactic halo
\citep{2010AIPC.1273..197W} point to a significant lack of post-AGB
stars in comparison with the prediction of older stellar evolution
models. The fastest evolution of our models can help to diminish these
discrepancies, but they  will hardly solve them. Because of the age of the
hosting population, the post-AGB stars in those systems should be of
low mass. In the case of the study of \cite{2010AIPC.1273..197W},
reproducing the number of post-AGB stars just by increasing the
evolutionary speed would require an evolution faster than that of the
low-mass models of \cite{2009A&A...508.1343W}. As our models evolve
somewhat slower than the models of \cite{2009A&A...508.1343W} it is
clear that our models cannot solve the discrepancy. We  can reach a similar conclusion  about the discrepancy reported by \cite{2008ApJ...682..319B}
in M32. M32 has an almost solar like metallicity and is composed of
two main populations: an intermediate age population (2-5 Gyr) of
stars that contribute to $\sim 40$\% of the mass and an old
population ($>5$ Gyr) that contributes to $\sim 55$\% of the mass
\citep{2012ApJ...745...97M}. This means that the vast majority of the
post-AGB stars should have progenitors between $\sim 1$ and $\sim 1.5$
$M_\odot$. While our models are faster than the models adopted by
\cite{2008ApJ...682..319B} for the same final mass, the difference is
smaller when compared at similar progenitor masses (this is also true
in the case of our M15 models). In the relevant mass range, our
post-AGB timescales are only $\sim 1.4$ and $\sim 2.6$ (for $M_i=$1
and 1.5 $M_\odot$, respectively)  times faster than the older
models. This seems insufficient to completely solve the discrepancy
with the observations.

Finally, a more quantitative comparison with the lifetimes derived by
\cite{2007A&A...462..237G} from the clusters of the Magellanic Clouds
point to some shortcomings of the present models.  While our models
reproduce the right qualitative behavior of the C and M star
lifetimes,  they quantitatively predict lower timescales  by
  a factor of a few (Fig. \ref{Fig:LMC-SMC}). This is particularly
significant at the higher metallicities of the LMC, where our models
predict timescales that are  more than four times too short for
both M- and C-type stars.  While the failure to quantitatively
reproduce the timescales of carbon stars might point to an
overestimation of the mass-loss rate, the duration of the M-type stars
phase cannot be solved this way. This is because in our sequences the
duration of the M-type star phase is determined by the intensity of
3DUP, which sets the number of thermal pulses after which the model
becomes a carbon star. Given that the CBM at the inner boundary of the
convective envelope is already set to a minimum in these sequences
($f_{\rm CE}=0$) a decrease of their dredge-up efficiency could only
be attained by a decrease in the CBM at the pulse-driven convective
zone ($f_{\rm PDCZ}$). However, as mentioned in the previous section,
a decrease of this value would lead to a disagreement between the O
abundances at the intershell and those observed in PG1159 stars.  It
might be necessary to explore alternative mixing processes and the
parameter space of pre-AGB stellar evolution models to reproduce all
AGB and post-AGB observables simultaneously.

\section{Conclusions}
\label{sec:conclusion}

We present a detailed grid of post-AGB sequences
computed with updated micro- and macrophysics. This is the first grid
of post-AGB sequences in the whole range of masses of interest for the
formation of PNe  that takes  the developments in stellar
astrophysics in the last 20 years into account.  The new models include updated
opacities both for the low- and high-temperature regimes and for the
C- and O-rich AGB stars. Also, conductive opacities and nuclear
reaction rates have been included according to the last developments
in the field. During the AGB phase, the models also include a
consistent treatment of the stellar winds for the C- and O-rich
regimes. In addition, they include  CBM processes during the thermal pulses and
previous evolutionary stages. This allows the new models to reproduce
several AGB and post-AGB observables not reproduced by the older grids
\citep{1994ApJS...92..125V,1995A&A...299..755B}.  In particular, the
new models reproduce the C/O ratios of several AGB and post-AGB
objects of the Galactic disk
\citep{1986ApJS...62..373L,1990ApJS...72..387S,1994MNRAS.271..257K,2009AstL...35..518M,2012A&A...543A..11M,2014ApJ...784..173D}
and also the mass range of carbon stars in the Magellanic Clouds
\citep{2007A&A...462..237G}.  In addition, the IFMR and intershell
abundances of the new post-AGB models are consistent with
semiempirical determinations  from globular clusters
\citep{2009ApJ...692.1013S} and the determinations in post-AGB
PG1159 stars \citep{Werner2016,2006PASP..118..183W}, respectively.

The main result from our study is that new post-AGB sequences predict
post-AGB evolutions, which are at least three to ten times faster than the
models of similar mass in the older grids
\citep{1994ApJS...92..125V,1995A&A...299..755B}.  Also, the new
post-AGB models are 0.1 to 0.3 dex more luminous than the older
sequences of similar remnant masses.  Qualitatively, the shorter
timescales are agreement with the results obtained by
\cite{2008PhDT.......290K} and \cite{2009A&A...508.1343W}, for a restricted mass
range and by means of a completely independent code but also assuming
updated modeling. The faster evolution of our post-AGB models is in
very good quantitative agreement with the empirical determination of
\cite{2014A&A...566A..48G} from the study of CSPNe in the Galactic
bulge. These agreements give  us confidence in the main
conclusion of our work.  The differences between the new models and
the older grids is traced back to the update in microphysics and,
in particular, to a better modeling of 3DUP events during the
TP-AGB. This improved modeling changes the IMFR and final C content of the envelope of
post-AGB models, leading to shorter post-AGB
timescales.

The new grids should have a significant impact in studies of CSPNe.  On the
theoretical side, we expect the much faster evolution will have a significant
impact for the formation and evolution of PNe. Radiation-hydrodynamic
numerical simulations \citep{2014AN....335..378S,2014MNRAS.443.3486T} of the
formation of PNe based on new post-AGB stellar evolution models would be very
valuable.  The faster post-AGB evolution of the new models calls into question
the conventional wisdom that single stellar evolution cannot lead to the
formation of PNe in globular clusters
\citep{2013ApJ...769...10J,2015AJ....149..132B}.  In particular, the fastest
evolution of the new models should allow for the formation of low-mass CSPNe
($M\sim 0.55M_\odot$) in line with spectroscopic and asteroseismic
determinations \citep{2010A&ARv..18..471A, Werner2016}. In addition, the
brighter core mass-luminosity relation of the new models (see
Fig. \ref{Fig:Core_Lum}) will affect estimations of the masses of post-AGB
stars based on their luminosity; e.g. \citealt{2013ApJ...769...10J}. Finally, it would also be worthwhile to carry out another study of the implications of
the new timescales for the properties of the PNLF for different populations
\citep{2004A&A...423..995M}.

Our models might help to diminish the discrepancies between the predicted and
observed number of post-AGB objects in M32 and the Galactic Halo. A simple
estimation of the timescales indicates that they might be unable to completely
solve these discrepancies, but a detailed study is necessary.

The agreement of the new models with all the previously mentioned AGB
and post-AGB observational constraints along with the much updated
modeling of the stellar physics, strongly suggest that these models
should be preferred over the older models.  However, it would be
  unwise to assume the current models to be perfect. Our sequences
fail to quantitatively reproduce the number of M- and C-type AGB stars
as derived from the Magellanic Clouds. As discussed in the previous
section, it is not clear that it is possible to quantitatively
reproduce all AGB and post-AGB observational constraints
simultaneously within the present framework for  CBM processes on
the AGB. An exhaustive calibration of models on the basis of the AGB
and post-AGB observables is needed to improve our knowledge
of the post-AGB phase.

Finally, but not of least importance, the sequences we present here were
computed under the assumption of steady winds on and after the
TP-AGB. While it is possible to devise tricks to avoid the
instabilities at the end of the TP-AGB, it is clear that there are
some physical reasons behind these instabilities \citep{1994A&A...290..807W,
  2012A&A...542A...1L} and a proper assessment of the consequences of
these instabilities is needed to improve our understanding of the
post-AGB phase. An almost total ejection of the envelope could
  produce post-AGB stars that depart from the AGB with less massive
  envelopes. Consequently, these instabilities could produce  much
  faster post-AGB evolutions than predicted by our current models.

The main grid of H-burning post-AGB sequences computed in this work,
as well as the M15 models (also presented in tables
\ref{tab:ZAMS-TPAGB-M15} and \ref{tab:pAGB-M15}), are provided in
tabulated electronic form and at similar points in the HR diagram to
allow for an easy interpolation. Sequences are available in electronic
form at at our web site {\tt http://www.fcaglp.unlp.edu.ar/evolgroup} and
at the CDS via anonymous ftp to {\tt cdsarc.u-strasbg.fr (130.79.128.5)} or
via {\tt http://cdsweb.u-strasbg.fr/cgi-bin/qcat?J/A+A/}.

\begin{acknowledgements}
  This work was supported by a fellowship for postdoctoral researchers
  from the Alexander von Humboldt Foundation. M3B thanks A. Weiss,
  A. Serenelli, L. Althaus, M. Viallet for helpful discussion during
  the development of this work and K. Werner for the
  new data on PG1159 stars.  A. Weiss is also warmly acknowledged for
  reading and commenting on the manuscript. We thank the first
  anonymous referee for very detailed reports and, in particular, the
  second referee (J. Lattanzio) for very constructive and helpful
  reports that strongly improved the final version of this manuscript.
  Finally, we are especially grateful to the editor (R. Kotak) for her
  patience and positive approach to the peer review process.
\end{acknowledgements}

\bibliographystyle{aa}


\Online
\begin{appendix} 

\section{ The physics behind the crossing time ($\tau_{\rm cross}$)}
\label{app:physics}
In sections \ref{sec:post-AGB} and \ref{sec:discussion} we have
discussed that the timescales of the post-AGB evolution ($
\tau_{cross}$) predicted by the new models are much  shorter than those
of the older sequences of \cite{1994ApJS...92..125V} and
\cite{1995A&A...299..755B} (Fig. \ref{Fig:CrossTimes}); this is in agreement
with the more  recent  results presented by \cite{2009A&A...508.1343W}
and \cite{2015ASPC..493...83M}. The value of $ \tau_{cross}$ during
the post-AGB phase is mainly set by the speed at which the remaining
H-rich envelope ($M_{\rm env}$) is consumed to its final value in the
white dwarf phase. Models that depart from the AGB with less massive
envelopes or with higher luminosities (i.e., faster H-burning) display shorter timescales. In what follows, we show that
updated microphysics and pre-AGB modeling,  together with carbon
  pollution of the envelope and lower values of $M_f(M_i)$, lead to
post-AGB models that are brighter and depart from the AGB with smaller
$M_{\rm env}$, leading to a faster post-AGB evolution.

 \begin{figure}
   \centering
   \includegraphics[width=\hsize]{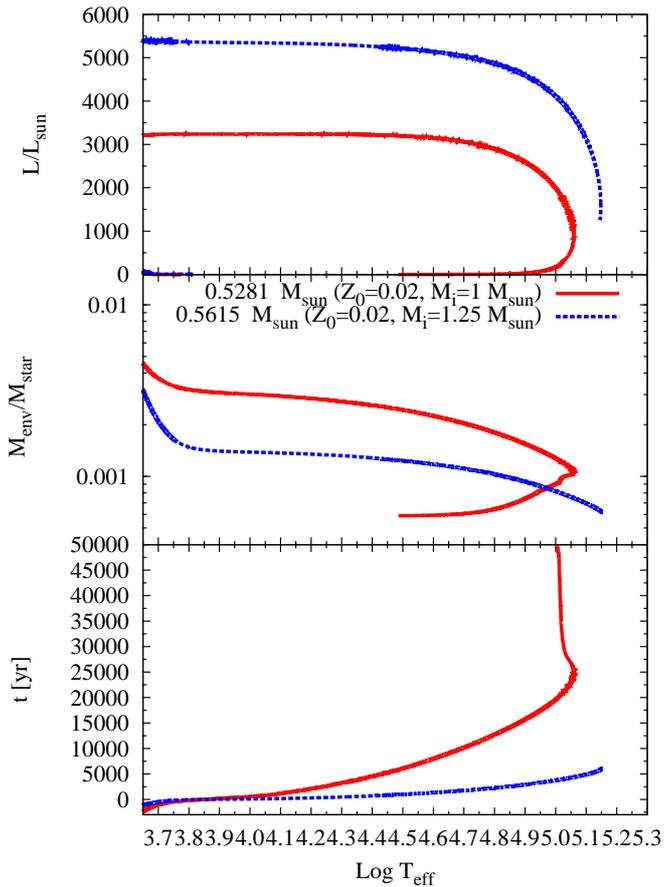}
   \caption{Evolution of the main quantities of the sequences with $M_f=0.5281
     M_\odot$ and $M_f=0.5615 M_\odot$ ($Z_0=0.02$).  These sequences have higher
     luminosities and smaller final envelope masses than the models of similar
     final mass of \cite{1983ApJ...272..708S}.}
         \label{Fig:M_env}
   \end{figure}
   Fig. \ref{Fig:M_env} shows the evolution of the main quantities of the
   sequences with $M_f=0.5281 M_\odot$ and $M_f=0.5615 M_\odot$ ($Z_0=0.02$)
   we present here.  These sequences do not undergo 3DUP on
   the TP-AGB and their envelope composition is not enriched in
   CNO elements. As a result of the similar masses and metallicities, these two
   sequences can be directly compared with the 0.546 $M_\odot$ and 0.565
   $M_\odot$ sequences of \cite{1983ApJ...272..708S}. The 0.546 $M_\odot$
   (0.565 $M_\odot$) of \cite{1983ApJ...272..708S} depart from the AGB
   (i.e., $\log T_{\rm eff}=3.85$) with $M_{\rm env}\sim 0.0073 M_\star$
   ($M_{\rm env}\sim 0.0022 M_\star$) and a post-AGB luminosity of $\sim 1500
   L_\odot$ ($\sim 3900 L_\odot$). As shown in Fig. \ref{Fig:M_env}, our
   $M_f=0.5281 M_\odot$ ($M_f=0.5615 M_\odot$) sequence departs from the AGB
   with $M_{\rm env}\sim 0.0031 M_\star$ ($M_{\rm env}\sim 0.0014 M_\star$)
   and a post-AGB luminosity of $\sim 3260 L_\odot$ ($\sim 5360 L_\odot$). We
   see that models with an updated treatment of the previous evolution, in
   particular, opacities and nuclear reaction rates, lead to models that depart
   from the AGB with smaller H-envelope masses and brighter luminosities.
    Consequently, the new models do evolve much faster than the older models. 

   There is another improvement of the new models that leads to even
   shorter timescales. As we discussed in Section \ref{sec:ZAMS-AGB},
   the new models are able to  agree with the expected carbon
   enrichment as observed in many AGB and post-AGB observables
   (Figs. \ref{Fig:LMC-SMC} and \ref{Fig:Comp-AGB-obs}). This is a
   consequence of the more  efficient 3DUP present in the new models.  Next, we discuss two numerical experiments to show
   that more  efficient 3DUP on the TP-AGB leads to post-AGB models that
   evolve faster than in the absence of 3DUP.

  \begin{figure}
   \centering
   \includegraphics[width=\hsize]{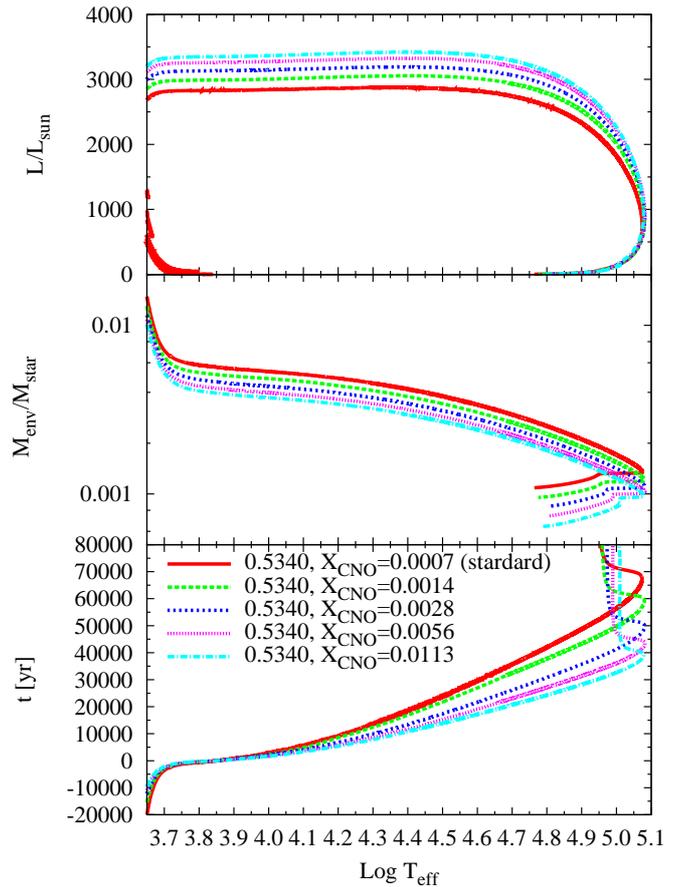}
   \caption{Evolution of the main quantities of the sequences with $M_f=0.5340
     M_\odot$ ($Z_0=0.001$) with different degrees of CNO pollution in the
     H envelope (see text for a discussion). There are higher luminosities,
     smaller final envelope masses and, consequently, shorter post-AGB
     timescales when the envelope is polluted with C and O.}
         \label{Fig:Experimento_CNO}
   \end{figure}
   Fig. \ref{Fig:Experimento_CNO} shows the main properties of a   model with $M_f=0.5340 M_\odot$ ($Z_0=0.001$) when different
   degrees of CNO pollution are added to the H-rich envelope. The
   original sequence corresponds to the $M_f=0.5340 M_\odot$
   ($Z_0=0.001$) of initially $0.9 M_\odot$ (see table \ref{tab:pAGB}), which
    does not undergo 3DUP episodes and has a CNO element mass
   fraction of $X_{\rm CNO}=0.0007$. For the other sequences in Fig.
   \ref{Fig:Experimento_CNO}, we artificially increased the CNO
   mass fraction by factors of 2, 4, 8. and 16, before the departure
   from the AGB, and recomputed the post-AGB evolution. The proportion
   of the polluting material is 82\% $^{12}$C and 12\% $^{16}$O so
   that  $N_{\rm C}/N_{\rm O}\sim 5$ for highest adopted pollution, as observed in the detailed models; see table \ref{tab:pAGB}.
   As shown in Fig. \ref{Fig:Experimento_CNO}, the higher the CNO
   pollution of the envelope, the higher the luminosity and the lower
   the mass of the H envelope at the departure from the AGB. A
   qualitative explanation of this behavior can be obtained from the
   study of toy envelope models; see \citep{1983ApJ...268..356T,
     1999A&A...351..161M}. The enrichment of the envelope with CNO
   element alters both the CNO burning rate and the molecular weight
   of the envelope, leading to higher luminosities for the same core
   mass. Models with efficient 3DUP consequently show shorter post-AGB
   timescales.

   Our models of $M_i\gtrsim 1.5 M_\odot$ undergo efficient 3DUP
   and CNO pollution of the envelope. Consequently, they show higher
   luminosities, smaller post-AGB H envelopes, and shorter post-AGB timescales
   than models with no efficient 3DUP (as those of the older
   grids). This effect is more important for sequences at low initial
   metallicities ($Z_0$), as the relative increase of the CNO elements is much
   higher.

  \begin{figure}
   \centering
   \includegraphics[width=\hsize]{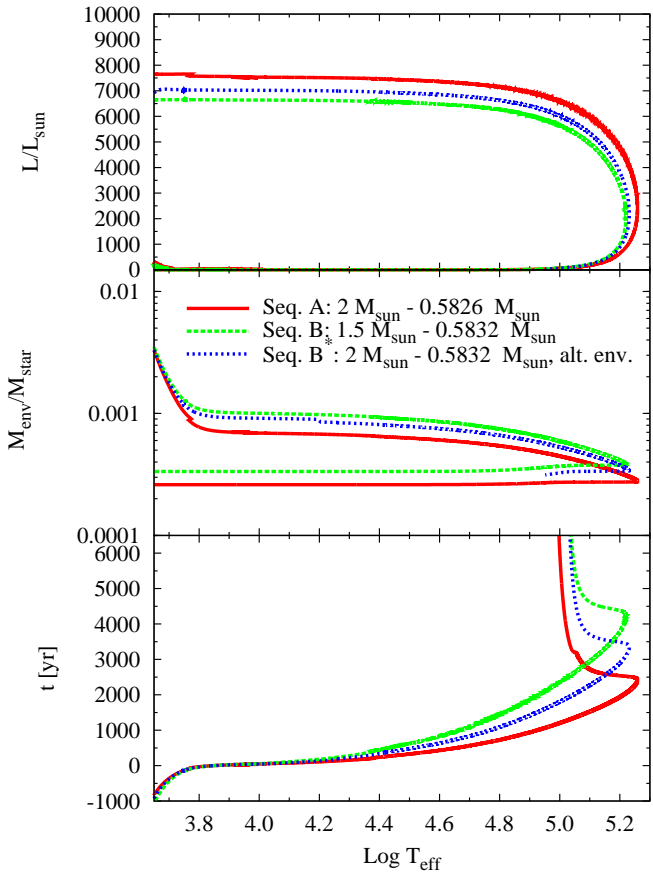}
   \caption{Evolution of the main quantities of the sequences A, B and B$^{*}$
   from the numerical experiment described in Section \ref{sec:discussion} and
 Table \ref{tab:experimento}.}
         \label{Fig:Experimento}
   \end{figure}
\begin{table}
\caption{Properties of the models of Fig. \ref{Fig:Experimento} at $\log T_{\rm eff}=5$.}            
\label{tab:experimento}     
\centering                     
\begin{tabular}{c c c c c c c}     
\hline\hline               
ID & Mass & $m_{\rm shell}$ &  $g_{\rm shell}$ & $L_{\rm surf}$& $ X^{\rm env}_{\rm C+N+O} $ & $\mu_{\rm env}$ \\  
  &$[M_\odot]$ & $[M_\odot]$ &  $[g_\odot]$ & $[L_\odot]$&   &   \\  

\hline                     
A  &  0.5826       & 0.58237 & 1248.6 & 6609 & 0.0197 & 0.624 \\   
B$^{*}$ &  0.5832   & 0.58289 & 1120.1 & 6026 & 0.0197 & 0.624 \\   
B  &  0.5832       & 0.58293 & 1135.7 & 5692 & 0.0121 & 0.616 \\   
\hline                                   
\end{tabular}
\end{table}

In addition to the CNO pollution of the envelope, 3DUP changes the
IFMR of the models; see \citealt{2009ApJ...692.1013S} for a
discussion. Intense 3DUP delays the growth of the HFC during the
TP-AGB evolution, leading to smaller final masses $M_f$ than in the
absence of 3DUP for the same initial mass $M_i$. Different HFC
histories lead to different core mass-luminosity relations and also to
different H envelopes at the departure from the AGB.  The following
simple numerical experiment shows this point. We have taken two of our
H-burning post-AGB sequences with different evolutionary histories, but
similar core masses, and changed the envelope on top of one of them
 to match the composition of the other;  $Z_0=0.01$ sequences
with initial masses $M_i=2$ and 1.5 $M_\odot$ in table \ref{tab:pAGB},
from now on sequences A and B, respectively. This was done on the
 TP-AGB to allow the H-shell to advance to the location of the
new composition and to give time for the envelope structure to
relax. As shown in Fig. \ref{Fig:Experimento}, when the envelope
composition of sequence A is put on top of the HFC of sequence B (from
now on sequence B$^{*}$), the luminosity is increased due to a higher
molecular weight and  the mass fraction of CNO elements. Yet, the luminosity
of sequence B$^{*}$ is lower than that of sequence A, in spite of
having the same envelope composition and almost the same core mass. In fact, the core mass is even slightly larger; see Fig. \ref{Fig:Experimento}. This
experiment shows that the influence of the core goes beyond the value
of its core mass.  The burning shell is not influenced either by the
  core mass or the core radius alone, but only through the
gravitational acceleration $g(r)$ at the location of the burning
shell.  Table \ref{tab:experimento} shows some relevant quantities of
the sequences A,B and B$^{*}$ at a similar point in the HR diagram.
As can be seen, the change in the envelope composition of $\sim$39\%
in the CNO-mass fraction\footnote{In the model with the altered
  envelope, the shell adjusts to a point with a slightly lower
  temperature. Because of the extremely high sensitivity of the CNO cycle
  to temperature, this prevents this change of $\sim$39\% to translate
  directly into the luminosity.} and of 1.3\% in the molecular weight
can help to understand the $\sim 5.5$\% change in the luminosity
between sequences with the same core but different envelopes (see
\citealt{1999A&A...351..161M} for a very similar discussion). However,
in spite of the $\sim 5.5$\% luminosity increase resulting from a different
envelope composition, model B$^{*}$ is still $\sim 9$\% less luminous
than sequence $A$ (whose core is 0.09\% smaller). A closer examination
of the model shows that the gravitational pull at the location of the
H-burning shell is $\sim 10$\% higher than that of sequence B and
B$^{*}$.  This experiment shows that at the moment of the departure
from the AGB our models show a core still warm enough that the
strict core mass-radius relation does not hold (they are yet not
``converged'' in the sense discussed by \citealt{1970AcA....20...47P})
. Consequently, previous evolutionary histories still play a
significant role in determining the post-AGB timescales and
luminosities. This also helps to explain why our $2 M_\odot$ ($Z_0=0.02$)
is $\sim 50$\% faster than our $1.5 M_\odot$ ($Z_0=0.02$) in spite of
having similar core masses and the second having almost twice the mass
fraction of CNO elements (table \ref{tab:pAGB}).  The influence of the
core through the gravitational pull at the H-burning shell is even
more important than the chemical change due to 3DUP.

We have shown that the short post-AGB timescales of the new post-AGB
models we present  (table \ref{tab:pAGB}) can be traced to
three main causes. First, as a result of the update in the micro- and
macrophysics of the evolutionary models, the new models depart with
smaller H envelopes and brighter luminosities.  Second, because of the CNO
pollution of the envelope produced by 3DUP in most of the sequences,
the post-AGB models have even smaller H envelopes  with brighter
 luminosities. Third, also because of the existence of 3DUP, the
post-AGB models of a given mass $M_f$ originate from initial models of
higher mass $M_i$. Post-AGB models derived from initial models of
higher mass have more compact HFCs and also depart from the AGB with
smaller H envelopes and with brighter  luminosities.

\section{M15 Post-AGB sequences \citep{2015ASPC..493...83M}}
\label{app:M15}
\begin{figure*}
            \includegraphics[width=\textwidth,clip,clip]{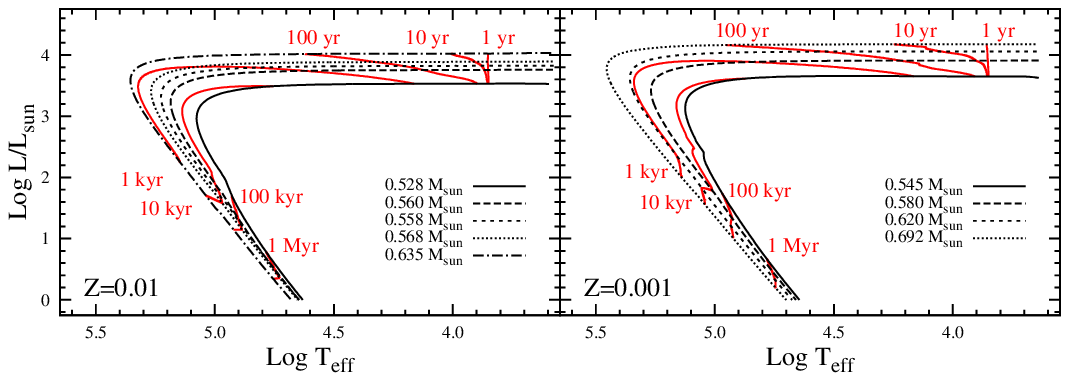}
            \caption{HR diagrams
      of the H-burning post-AGB sequences computed by M15 for different
      masses and initial metallicities (Table \ref{tab:pAGB-M15}). Tracks are presented from the
      beginning of the post-AGB phase when the H-rich envelope drops below
      $M_{\rm env}=0.01 M_\star$ to the moment in which the star has already
      entered its white dwarf cooling sequence at $L_\star=L_\odot$. At that
      point gravitational settling should have already started to turn
      post-AGB stars into DA-WDs, a process not included in the present
      computations. Red lines indicate computed isochrones for different ages
      since the zero point defined at $\log T_{\rm eff}=3.85$. }
         \label{Fig:pAGB-M15}
   \end{figure*}

M15 models have been computed with the same code and microphysics but a
different choice of the  CBM processes during the TP-AGB. In
particular, M15 models were computed under the assumption of efficient  CBM process at the base of all convective envelopes,
setting\footnote{There are two other minor differences with the main grid of
  models we present: the wind limit of eq. \ref{eq:SSLim} was not
  applied and the $Z_0$ dependence in the hot winds (eq. \ref{eq:cspn}) was
  not included. Neither of these two differences play a dominant role in the
  final post-AGB timescales.} $f_{\rm CE}=0.13$. Consequently, M15 models
undergo more efficient 3DUP episodes than the models of our main
grid; tables \ref{tab:pAGB} and \ref{tab:ZAMS-TPAGB}. As discussed in Section
\ref{sec:discussion}, previous evolutionary history can play an important role
in the determination of the post-AGB timescales of the models. In particular,
models with more efficient 3DUP during the TP-AGB have
lower final masses ($M_f$) for the same initial mass ($M_i$). Consequently,
and in line with the discussion of sections \ref{sec:post-AGB} and
\ref{sec:discussion}, more efficient 3DUP leads to faster post-AGB
evolution for the same value of $M_f$. While M15 models fail to reproduce the
IFMR at almost solar metallicities, we consider these models valuable at
assessing the uncertainty behind the results we present. In fact,
these models predict lifetimes of the C-rich phase in better agreement with the
determinations of \cite{2007A&A...462..237G} from the study of the Magellanic
Clouds. For this reason, we present in this appendix the main properties of the
M15 model in the previous evolutionary history (table
\ref{tab:ZAMS-TPAGB-M15}) and during the post-AGB phase (table
\ref{tab:pAGB-M15} and Fig. \ref{Fig:pAGB-M15}).

\begin{table*}
\caption{Main properties of the M15 sequences from the ZAMS to the TP-AGB.}
\label{tab:ZAMS-TPAGB-M15}
\centering
\begin{tabular}{cccccccccccc} 
\hline\hline             
  $M_i$ & $\tau_{MS}$ & $\tau_{RGB}$ & HeCF & $\tau_{HeCB}$  &  $\tau_{eAGB}$ & $M_c^{1TP}$ &  $\tau_{TP-AGB(M)}$& $\tau_{TP-AGB(C)}$& \#TP & $M_f$ & $N_{\rm C}/N_{\rm O}$\\
  $[M_\odot]$ & [Myr] &[Myr]   &  & [Myr]       &  [Myr]     &$[M_\odot]$& [Myr]             & [Myr]&     (AGB)       &      $[M_\odot]$  & \\
\hline
  \multicolumn{12}{c}{$Z_0=0.01$}\\ 
\hline
   1.00     &   8539.1     &   1847.8  &  yes        &   112.45     &   10.825     &   0.5119     &  0.71950     &   0.0000     &  4 &   0.5282     & .404     \\
  1.50     &   1989.7     &   481.76   &  yes       &   102.81     &   9.6305     &   0.5271     &  0.81428     &  0.26368     &  7 &   0.5595     & 1.77     \\
  2.00     &   969.87     &   59.771   &  no       &   162.42     &   13.899     &   0.5062     &   1.5132     &  0.65170     & 12 &   0.5584     & 2.52     \\
  2.50     &   534.59     &   9.2022   &  no       &   165.72     &   10.055     &   0.5434     &  0.61113     &   1.1034     & 11 &   0.5678     & 3.80     \\
  3.00     &   334.09     &   4.2506   &  no       &   87.321     &   4.8755     &   0.6386     &  0.17697     &  0.34383     &  8 &   0.6352     & 3.16     \\
  4.00     &   164.01     &   1.4422   &  no       &   27.686     &   1.6394     &   0.8018     &  0.29503E-01 &  0.10702     & 11 &   0.7968     & 1.95     \\
\hline
  \multicolumn{12}{c}{$Z_0=0.001$}\\ 
\hline
 1.00     &   5649.1     &   848.89   &  yes        &   86.936     &   8.5632     &   0.5240     &  0.62546     &  0.24255     &  4 &   0.5451     & 8.44     \\
  1.50     &   1342.9     &   319.05  &  yes        &   84.146     &   7.3278     &   0.5564     &  0.52748E-03 &   1.0151     &  7 &   0.5804     & 7.78     \\
  2.00     &   616.85     &   15.556  &  no         &   158.36     &   5.6971     &   0.6167     &  0.82512E-01 &  0.77219     & 10 &   0.6202     & 8.29     \\
  2.50     &   399.46     &   6.0777  &  no         &   82.517     &   3.3881     &   0.7021     &  0.20568E-03 &  0.56845     & 12 &   0.6916     & 7.03     \\
\hline\\
\multicolumn{12}{p\textwidth}{ $M_i$: Initial mass of the model (at ZAMS).  $\tau_{MS}$:
    Duration of the main sequence, until $X^{\rm center}_{\rm
      H}=10^{-6}$. $\tau_{RGB}$: Lifetime from the end of the main
    sequence to He-ignition set at $\log L_{\rm He}/L_\odot=1$.
    HeCF: Full He-core flash (and subflashes) at the beginning of the
    core He-burning phase. $\tau_{HeCB}$: Lifetime of core He-burning,
    until $X^{\rm center}_{\rm He}=10^{-6}$.  $\tau_{eAGB}$: lifetime
    of the early AGB phase from the end of core helium burning to the
    first thermal pulse. $M_c^{1TP}$: Mass of the H-free core at the
    first thermal pulse (defined as those regions with $X_{\rm
      H}<10^{-4}$). $\tau_{TP-AGB(M)}$: Lifetime of the star in the
    TP-AGB as a M-type star ($N_{\rm C}/N_{\rm O}<1$). $\tau_{TP-AGB(C)}$: Lifetime of
    the star in the TP-AGB as a carbon star ($N_{\rm C}/N_{\rm O}>1$). \#TP: Number of
    thermal pulses on the AGB. $M_f$: Final mass of the
    star. $N_{\rm C}/N_{\rm O}:$ C/O ratio in number fraction at the end of
    the TP-AGB phase.}
\end{tabular}
\end{table*}

\begin{table*}
\caption{ Main post-AGB properties of the H-burning
      sequences computed by M15.}
\label{tab:pAGB-M15}
\centering
\begin{tabular}{cccccccccc} 
\hline\hline   
  $M_i$ & $M_f$& $\tau_{tr}$ & $ \tau_{cross}$   &  $X_{\rm H}$ &  $X_{\rm He}$ &  $X_{\rm C}$ &  $X_{\rm N}$ &  $X_{\rm O}$ & $\Delta M_{\rm env}^{\rm winds}/\Delta M_{\rm env}^{\rm total}$ \\
  $[M_\odot]$ & $[M_\odot]$ & [kyr]  &  [kyr]   &  &   & &   &   &  \\
\hline
  \multicolumn{10}{c}{$Z_0=0.01$}\\ 
\hline
 1.00     &   0.5282     & 8.54     & 22.9     & .717     &.273     &.145E-02 &.876E-03 &.477E-02 &0.177    \\
  1.50     &   0.5595     & 3.69     & 5.38     & .706     &.275     &.842E-02 &.109E-02 &.632E-02 &0.228     \\
  2.00     &   0.5584     & 2.68     & 2.96     & .681     &.288     &.165E-01 &.151E-02 &.875E-02 &0.234     \\
  2.50     &   0.5678     & 2.03     & 1.74     & .661     &.302     &.226E-01 &.166E-02 &.793E-02 &0.237     \\
  3.00     &   0.6352     & 1.39     & .717     & .675     &.299     &.146E-01 &.177E-02 &.616E-02 &0.308     \\
  4.00     &   0.7968     & .934     & .714E-01 & .662     &.313     &.876E-02 &.598E-02 &.589E-02 &0.551     \\
\hline
  \multicolumn{10}{c}{$Z_0=0.001$}\\ 
\hline     
 1.00     &   0.5451     & 6.26     & 16.2     & .669     &.294     &.310E-01 &.400E-03 &.490E-02 &0.203     \\
  1.50     &   0.5804     & 2.33     & 2.47     & .676     &.286     &.318E-01 &.176E-03 &.544E-02 &0.265     \\
  2.00     &   0.6202     & 1.36     & .806     & .615     &.321     &.503E-01 &.197E-03 &.809E-02 &0.285     \\
  2.50     &   0.6916     & 1.03     & .312     & .664     &.290     &.354E-01 &.250E-03 &.670E-02 &0.333     \\
\hline
\multicolumn{10}{p\textwidth}{ $M_i$: initial mass of the model (at ZAMS). $M_f$:
      final mass of the star. $\tau_{\rm tr}$: timescale from the end
      of the AGB (taken at $M_{\rm env}=0.01 M_\star$) to the moment
      in which $\log T_{\rm eff}= 3.85$. $ \tau_{cross}:$ timescale
      from the moment in which $\log T_{\rm eff}= 3.85$ to the point
      of maximum effective temperature. $X_{\rm H}$, $X_{\rm He}$,
      $X_{\rm C}$, $X_{\rm N}$, and $X_{\rm O}$: H, He, C, N and O surface abundances of
      the post-AGB models. $\Delta M_{\rm env}^{\rm winds}$ and $\Delta
      M_{\rm env}^{\rm total}$: reduction of the H-rich envelope
      ($M_{\rm env}$), from $\log T_{\rm eff}= 3.85$ to the point of maximum
  $T_{\rm eff}$, due to winds and due to the combined effect of winds
  and H-burning, respectively.}
\end{tabular}
\end{table*}

\section{Numerical improvements and convergence issues}
\label{app:numerics}
The computation of the very end of the TP-AGB  suffers from 
convergence problems
\citep{1994A&A...286..121W,2009A&A...508.1343W,2012A&A...542A...1L}. This implies that a lot of human time (babysitting) is required to
compute the transition from the TP-AGB to the CSPNe phase. Even when
codes converge, convergence happens at the expense of prohibitively
small time steps (even down to $\Delta t \sim 1 $ hour).  In the
following, we describe the methods and tricks adopted to improve the
convergence of the code during the very end of the AGB.

The main difference with previous works is that now remeshing in LPCODE can be
done by checking the assumption of linearity  made on the differential
equations of structure, $$dy_i/dx=f_i(y_1,y_2,y_3,y_4),$$ as suggested by
\cite{1994A&A...286..121W}, where $y_i$ are some function of the structure
variables $l$, $P$, $r,$ and $T$ and $x$ is a function of the Lagrangian
coordinate $m$. However, we adopt a much more straightforward approach than
that of \cite{1994A&A...286..121W}. Instead, we simply check that the relative
changes of $dy_i/dx$ from one shell ($n$) to the next ($n+1$) are kept under a
certain value, i.e.,  $$|2(f^{n}_i-f^{n+1}_i)/(f^{n}_i+f^{n+1}_i+\epsilon)|<
\delta_i,$$ where $\epsilon$ is an arbitrary value to avoid divergence in the
case of $f^{n+1}_i\sim f^{n}_i\sim 0$. According to this criterion, a mesh point is
added when it is not fulfilled at least for one equation, and removed when all
equations fulfill this mesh point by more than one order of magnitude.

Despite numerical improvements and debugging, our computations of the
end of the AGB still face several converge problems. In particular, two
main instabilities were found during the computation of these
sequences. Although we have not studied them in detail they can be
traced back to the instabilities already mentioned in the
literature. Both were found in late TP-AGB models in which the H-rich
envelope mass has been significantly reduced and is already comparable
to the mass of the HFC. The first instability might already develop in
intermediate-mass models  during the TP-AGB just after the
thermal pulses when the envelope expands as a consequence of the
energy injected by the thermal pulse and the H-burning shell is
temporarily shut down. We find that as the luminosity of the star
increases a behavior arises that is similar to the thermal
hydrogen instability described by \cite{1994A&A...290..807W}, 
and our simulations crash. While it would be worth studying whether
such instability finally develops into a sudden ejection of the upper
parts of the envelope, our code is unable to deal with such
hydrodynamic situations. The instability develops in the upper part of
the H-rich envelope and  thus it is very unlikely that it would
lead to the ejection of the whole H-rich envelope and the termination
of the AGB phase.  We assume that the mean mass
ejection during the TP-AGB is well described by the steady
winds (eqs.  \ref{eq:sk}, \ref{eq:Mdot-MAGB}, \ref{eq:Mdot-CAGB},
\ref{eq:SSLim} and \ref{eq:cspn}). To avoid the lack of
convergence, we force the integration of the outer stellar envelope
(were $dS/dt=0$ is assumed; see \citealt{2003A&A...404..593A} for
details) below the point where H recombination takes place. A second
instability was found in intermediate-mass ($M_i \gtrsim 2 M_\odot$)
models already departing from the AGB at the base of the H-rich
envelope where the iron opacity peak is found, causing convection to
become slightly superadiabatic and radiation pressure to become
dominant, at some points becoming even more than one order of
magnitude larger than the gas pressure. This instability is thus
related to the instability discussed in detail by \cite{2012A&A...542A...1L}
and references therein.  In our sequences we do not find a clear
runaway instability as in the study of \cite{2012A&A...542A...1L},
probably because of the lower luminosity and mass of our sequences. This
instability does, however, force the timestep to very small values
(even less than one hour).  This transition time, between
the end of the AGB and the beginning of the CSPNe phase, is on the
order of 1000 to 10000 yrs and, with such small timesteps, it would
require several millions of timesteps to compute this short-lived
phase. This has to be compared with the few thousand to tens of 
  thousands of models required to compute the whole evolution up to
the beginning of the AGB phase. Even if the models do not fail to
converge, such small timesteps are unaffordable. This instability
becomes worse for more massive (i.e., luminous) models. In order to
overcome this difficulty we have forced the bottom of the convective
envelope to the adiabatic regime. As discussed by
\cite{2012A&A...542A...1L} forcing adiabatic convection helps to avoid
this instability. In fact, when this is carried out, we find that the
oscillations in the $e_g=-T \partial s/\partial t$ term in the energy
equation are damped, which  facilitates much larger timesteps. This was done only at the very bottom of the convective envelope
where convection is only slightly superadiabatic  to allow the
outer regions of the envelope to be superadiabatic and thus obtain the
correct stellar radius and effective temperature. This is important
because our mass-loss prescriptions (eqs.  \ref{eq:sk},
\ref{eq:Mdot-MAGB}, \ref{eq:Mdot-CAGB}, \ref{eq:SSLim} and
\ref{eq:cspn}) are strongly dependent on having the right surface
parameters. In addition we have checked that this procedure does not
affect our results by computing a $M_i = 2 M_\odot$ ($Z_0=0.01$) sequence
with and without this alteration of the temperature profile at the
base of the convective envelope and obtaining the same
results. Finally, a different convergence problem arose (only) in the
case of our most massive and  highest metallicity sequence
($M_i=4 M_\odot$, $Z_0=0.02$). In that sequence convergence problems
also arose during the integration of the outer boundary conditions of
the model at high effective temperatures ($\log T_{\rm eff}\gtrsim
5$). This is not surprising since, at those $T_{\rm eff}$ values, the
opacity bump of the iron group elements is located close to the
photosphere and even in the atmosphere. Then at high luminosities,
the Eddington limit is very close and hydrostatic solution might
not exist. We avoided this convergence issue by adopting a very rough
boundary condition for the photospheric pressure as
$P_{\tau=2/3}=2GM_\star/(3 R_\star^2 \bar{\kappa})$,  where
$\bar{\kappa}$ is a mean value of the Rosseland opacity in the
atmosphere.

\end{appendix}

\end{document}